\documentclass[preprint,preprintnumbers,superscriptaddress,nofootinbib,aps,10pt]{revtex4-1}
 \usepackage{amsmath,amssymb,bm,epsfig}
 \usepackage{color}
 \usepackage{natbib}
 \usepackage{hyperref}
 \usepackage{ulem}
 \usepackage{graphicx}
 \usepackage{xcolor,colortbl}
 \usepackage{pifont}

\def \beq{\begin{equation}}
\def \eeq{\end{equation}}
\def \beqa{\begin{eqnarray}}
\def \eeqa{\end{eqnarray}}
\def \la{\langle}
\def \ra{\rangle}
\def \l{\left(}
\def \r{\right)}

\usepackage{xspace}

\newcommand{\sNN}{\sqrt{s_{\rm NN}}}

\begin{document}

\title{Probing Pb+Pb collisions at $\sqrt{S_{NN}}=2760$ GeV with spectators}

\author{Vipul Bairathi}
\email{vipul.bairathi@niser.ac.in}
\affiliation{School of Physical Sciences, National Institute 
of Science Education and Research, Jatni, 752050, India}

\author{Sandeep Chatterjee}
\email{sandeepc@vecc.gov.in}
\affiliation{Theoretical Physics Division, 
Variable Energy Cyclotron Centre, 1/AF Bidhannagar, 
Kolkata, 700064, India}

\author{Md. Rihan Haque}
\email{rihan.h@niser.ac.in}
\affiliation{School of Physical Sciences, National Institute 
of Science Education and Research, Jatni, 752050, India}
  
\author{Bedangadas Mohanty}
\email{bedanga@niser.ac.in}
\affiliation{School of Physical Sciences, National Institute 
of Science Education and Research, Jatni, 752050, India}

\begin{abstract}
There is event by event geometric as well as quantum fluctuations in the initial condition 
of heavy-ion collisions. The standard technique of analysing heavy-ion collisions in bins 
of centrality obtained from final state multiplicity averages out the various initial 
configurations and thus restricts the study to only a limited range of initial conditions. 
In this paper, we propose an additional binning in terms of total spectator neutrons in an event. 
This offers us a key control parameter to probe events with broader range of initial 
conditions providing us an opportunity to peep into events with rarer initial conditions 
which otherwise get masked when analysed by centrality binning alone. We find that the 
inclusion of spectator binning allows one to vary $\varepsilon_2$ and $\varepsilon_3$ 
independently. We observe that the standard scaling relation 
between $\displaystyle{v_2/\varepsilon_2}$ and $\frac{1}{S}\frac{dN_{\text{ch}}}{d\eta}$ 
exhibited by centrality bins is strongly broken by the spectator neutron bins. 
Further, the acoustic scaling relation between $\displaystyle{\ln\l v_n/\varepsilon_n\r}$ 
and transverse system size is also broken- the strength of the breaking being sensitive 
to the binning procedure. The 
introduction of the spectator binning allows us to tune over a wide range viscosity driven 
effects for events with varying initial states but similar final state multiplicity.
\end{abstract}
\maketitle

\section{Introduction}
\label{sec.intro}
Of all the stages in a heavy-ion collision (HIC), the initial stage is the least understood. 
However, in order to perform a sensitive test of the theoretical framework, e.g. relativistic 
viscous hydrodynamics, that correctly describes the evolution of the strongly interacting matter 
produced in HIC experiments and therof allows unambiguous extraction of the medium properties, 
e.g. values of the transport coefficients, require a precise knowledge of the initial state 
(IS). It has been shown that depending on the choice of the initial condition that one chooses to 
evolve the relativistic hydrodynamic equations, the value of the extracted shear viscosity to 
entropy density ratio at the RHIC 200 GeV can vary by a factor of 2~\cite{Drescher:2007cd,
Romatschke:2007mq,Luzum:2008cw,Song:2010mg,Roy:2012jb,Roy:2012pn}. 

The nuclei used in HIC experiments are extended objects. This results in event by event (E/E) 
geometrical fluctuations in addition to the intrinsic quantum fluctuations of the nuclear wave 
function. The geometry of the nucleus ensures that various characteristics of the IS in HICs like the 
number of wounded nucleons $N_{\text{part}}$, number of binary collisions $N_{\text{coll}}$, shape 
of the overlap region, say the ellipticity $\varepsilon_2$, are all correlated with the impact parameter 
$b$. However, none of the above IS collision attributes are directly observed in experiments. This 
makes the job to constrain the IS very challenging. The standard method uses the final state (FS) 
charged particle multiplicity to characterize the events into different centrality classes corresponding 
to different ISs. However, the geometric and quantum E/E fluctuations in the 
IS result in appreciable variation of $b$, $N_{\text{part}}$, $N_{\text{coll}}$, $\varepsilon_2$ 
etc., even within the same centrality bin. Thus, a lack of proper knowledge of the IS is a major hindrance 
towards carrying out precise comparisons between theory and experiments. In this work, we focus on the 
spectators (those nulceons which do not participate in the collision) and show that it is possible 
to extract vital information of the IS by analysing them. The significant role played by spectator 
asymmetry in the various experimental observables and the possibility of selecting special initial 
configurations in HIC using deformed U nuclei has been pointed out recently~\cite{Chatterjee:2014sea, Bairathi:2015uba}. 
In a study based on Monte Carlo Glauber 
model simulations it was suggested that spectator asymmetry could be used to trigger specific collision 
configurations called Body-Tip with sufficient magnetic field and much lower ellipticity which can 
lead to the disentanglement of chiral magnetic effect from its dominant background anisotropic flow 
in $U+U$ collisions~\cite{Chatterjee:2014sea}. In Ref.~\cite{Bairathi:2015uba} it was demonstrated 
based on A Multi Phase Trasport (AMPT) model simulations that spectator asymmetry could be 
utilised to identify Body-Tip collision configurations in $U+U$ with large forward-backward 
asymmetry in particle production as well as considerably smaller elliptic flow, $v_2$. In this 
work, we look into the possibility of probing HICs of non-deformed nuclei using spectators. 
We will in particular focus on $Pb+Pb$ collisions at $\sNN=2760$ GeV. A recent study has pointed 
out the correlation between forward-backward asymmetry in particle production and spectator asymmetry 
in $Pb+Pb$ collisions at $\sNN=2760$ GeV~\cite{Jia:2015jga}. 
  
We have constructed an IS observable, namely the total spectator neutron (L+R), which is the sum of 
the left going (L) and right going (R) spectator neutrons that are detected by the zero degree calorimeters (ZDC). 
The spectator protons never reach the ZDC as they are bent by the 
magnetic field and hit the beam pipe wall in experiments say for example at RHIC. 
Hence we do not consider them. With the present design of the ZDC, faithful measurements of spectator 
neutrons are limited to central and mid-central collisions upto $\sim40\%$ centrality due to clustering effects~\cite{Abelev:2013qoq}.
However, there are suggestions for advanced designs with much improved ZDC performance which could be implemented 
in the future~\cite{Tarafdar:2014oua}.
In the absence of the IS fluctuations, $L+R$ and $b$ would have a one to one correspondence. Both quantities 
are zero for full overlap collisions and increase for peripheral collisions. The advantage of $L+R$ is that 
it is an observable measured in experiments while $b$ is never measured.
We demonstrate that by performing a further binning over $L+R$ in addition to the standard centrality binning, 
it is possible to probe the fireball with novel IS conditions as compared to centrality binning alone. This 
allows us to perform a more accurate comparison between model predictions and data, thus enabling a more 
precise modelling and comprehensive understanding of the response of the strongly interacting 
matter to the IS conditions created in HIC experiments. The outline of the paper is as follows: in the next 
section~\ref{sec.model} we discuss the details of the AMPT model and our simulation as well as
our proposed methodology to bin events. In Sec.~\ref{sec.result}, we show the main results obtained with this 
new binning procedure and finally summarise in Sec.~\ref{sec.summary}.

\section{Model}
\label{sec.model}

\begin{figure*}[htb]
 \begin{center}
  \scalebox{1}{
  \includegraphics[width=0.3\textwidth]{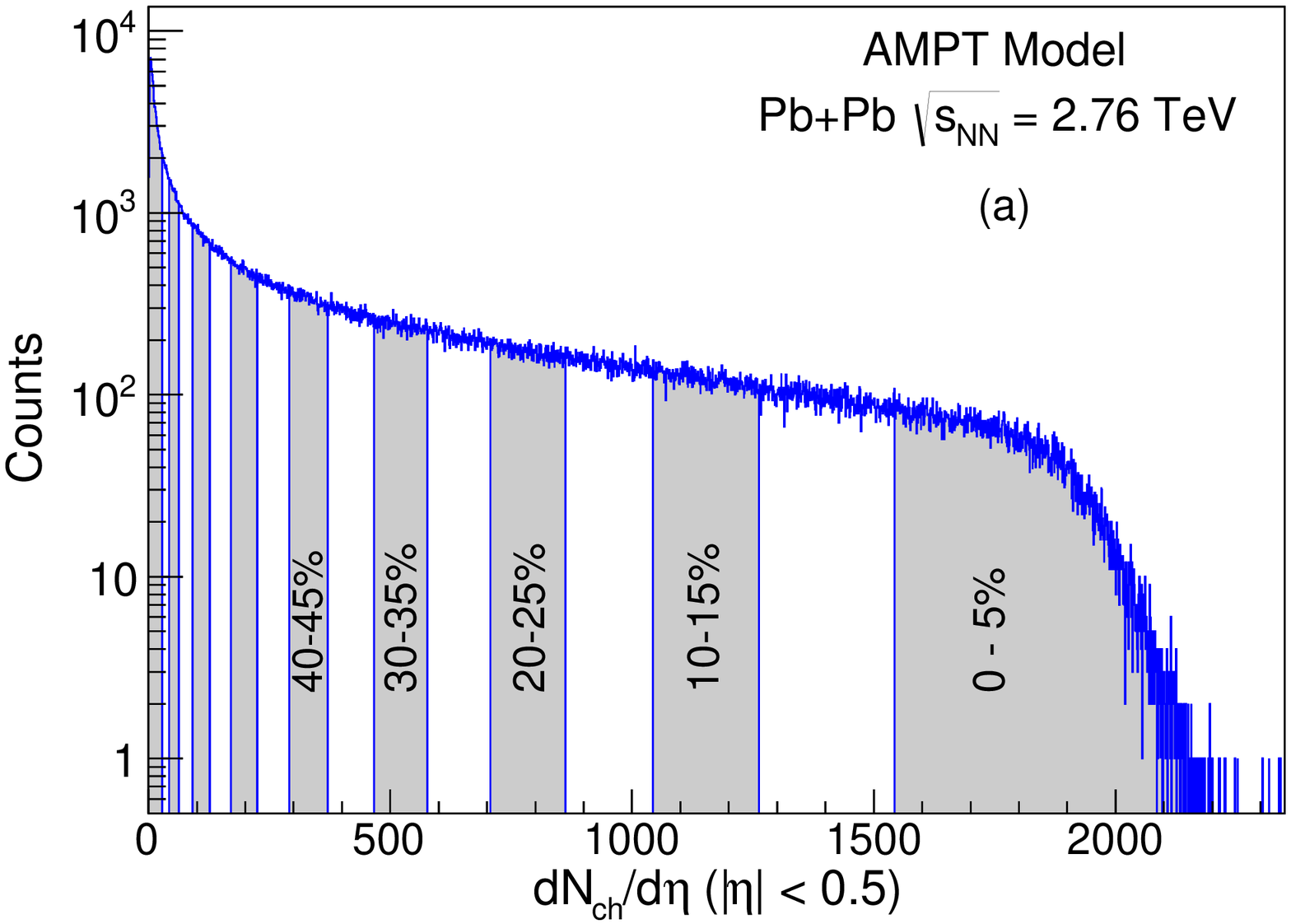}
  \includegraphics[width=0.3\textwidth]{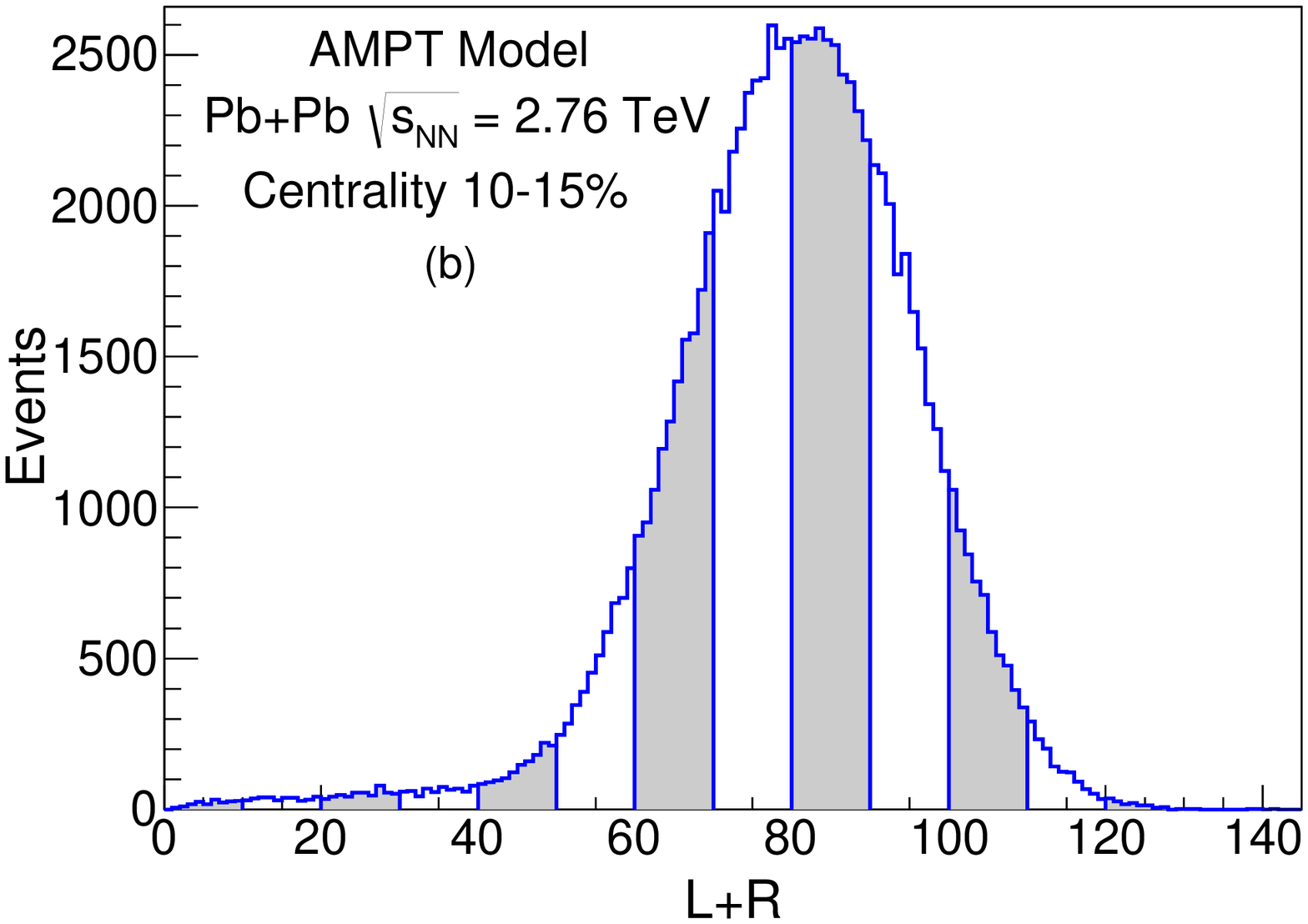}
  \includegraphics[width=0.3\textwidth]{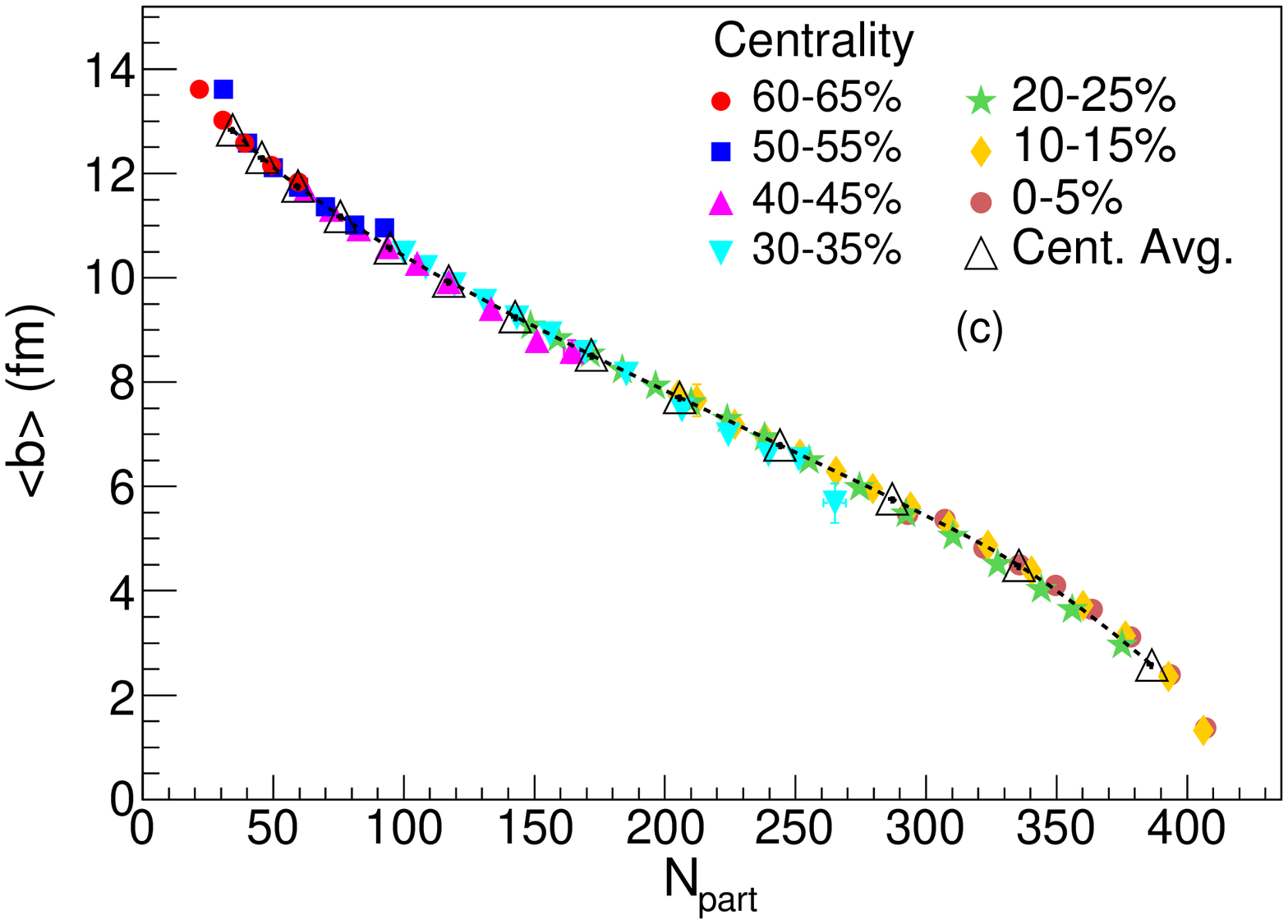}
  }  
\caption{(Color online) Left: The multiplicity distribution for min bias Pb-Pb events at $\sqrt{s_{NN}}=2760$ GeV. 
In alternate white and grey bands, different centrality bins are also shown. Middle: The total spectator neutron 
number $L+R$ distribution for the $\l10-15\r\%$ centrality. The different $L+R$ bins are also shown. Right: The impact 
parameter $b$ for the different different centrality and $L+R$ bins are shown.}
\label{fig.zdc}
\end{center}
\end{figure*}

We have simulated $Pb+Pb$ collisions at $\sNN=2760$ GeV using A Multi Phase Transport (AMPT)~\cite{Lin:2001zk,Lin:2004en} model
in the Default version. The AMPT model uses the same initial conditions as HIJING~\cite{Wang:1991hta}. Zhangs parton 
cascade follows to take into account partonic interactions~\cite{Zhang:1997ej} which finally recombine with their parent 
strings that fragment into hadrons within the Lund String Fragmentation model~\cite{Andersson:1983ia}. There is a final 
stage hadronic afterburner before the hadrons freezeout. In this study we have analysed $\sim2\times10^6$ events.

The standard practice is to first categorize events into different centrality classes according to charged particle 
multiplicity. This is shown in Fig.~\ref{fig.zdc} (a). The different centrality classes are shown in alternate white and 
grey bands. Here we propose to do a second round of binning with the observable $L+R$ within each 
centrality bin. The $L+R$ binning is illustrated in Fig.~\ref{fig.zdc} (b) where the $L+R$ distribution is shown for 
$\l10-15\r\%$ centrality. The $L+R$ distribution shows a prominent peak around 85-95 and falls off rapidly 
on either side - the number of events drop by a factor of 5 as $L+R$ shifts by $\sim20$. Thus when the analysis 
is performed based on centrality binning alone, we mainly study properties of the events with total spectator neutrons around 
85-95. Here with the introduction of the $L+R$ binning, we can investigate properties of the rare events with fewer or 
higher values of $L+R$ compared to the centrality mean value. This is the basic reason why additional $L+R$ binning 
on top of the centrality binning allows us to study new IS conditions in HICs. In Fig.~\ref{fig.zdc} (c) we have 
shown the variation of $b$ with $N_{part}$. We find the centrality and $L+R$ bins to follow the same trend. However as shown 
in Figs.~\ref{fig.IS}, \ref{fig.Geom} and \ref{fig.ncollgeom}, other IS attributes show different trends along centrality 
and $L+R$ bins that finally translate into different FS obsevable trends.

\section{Results and Discussion}
\label{sec.result}

\begin{figure*}[htb]
 \begin{center}
  \scalebox{1}{ 
  \includegraphics[width=0.45\textwidth]{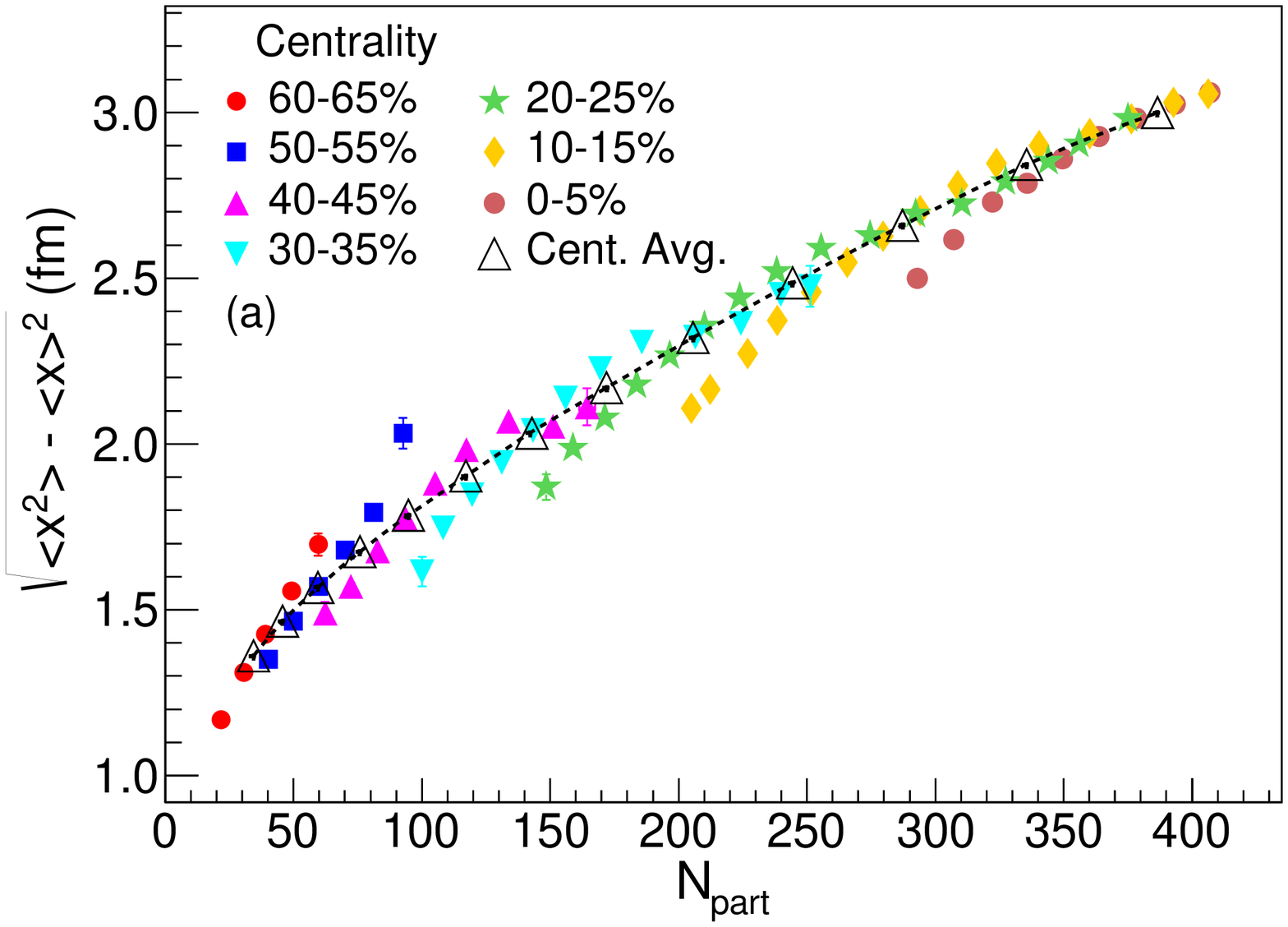}
  \includegraphics[width=0.45\textwidth]{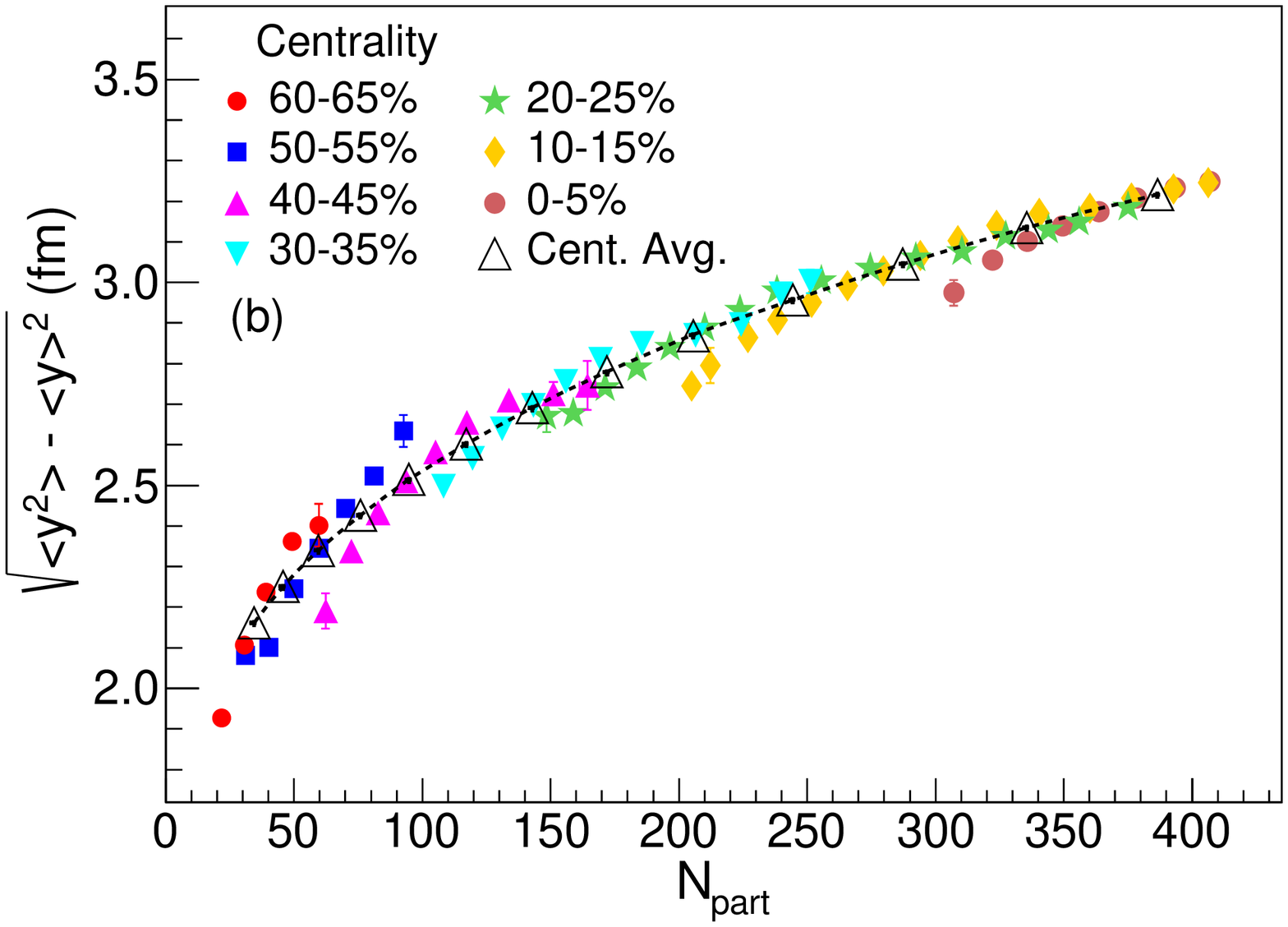}
  }  
  \scalebox{1}{
  \includegraphics[width=0.45\textwidth]{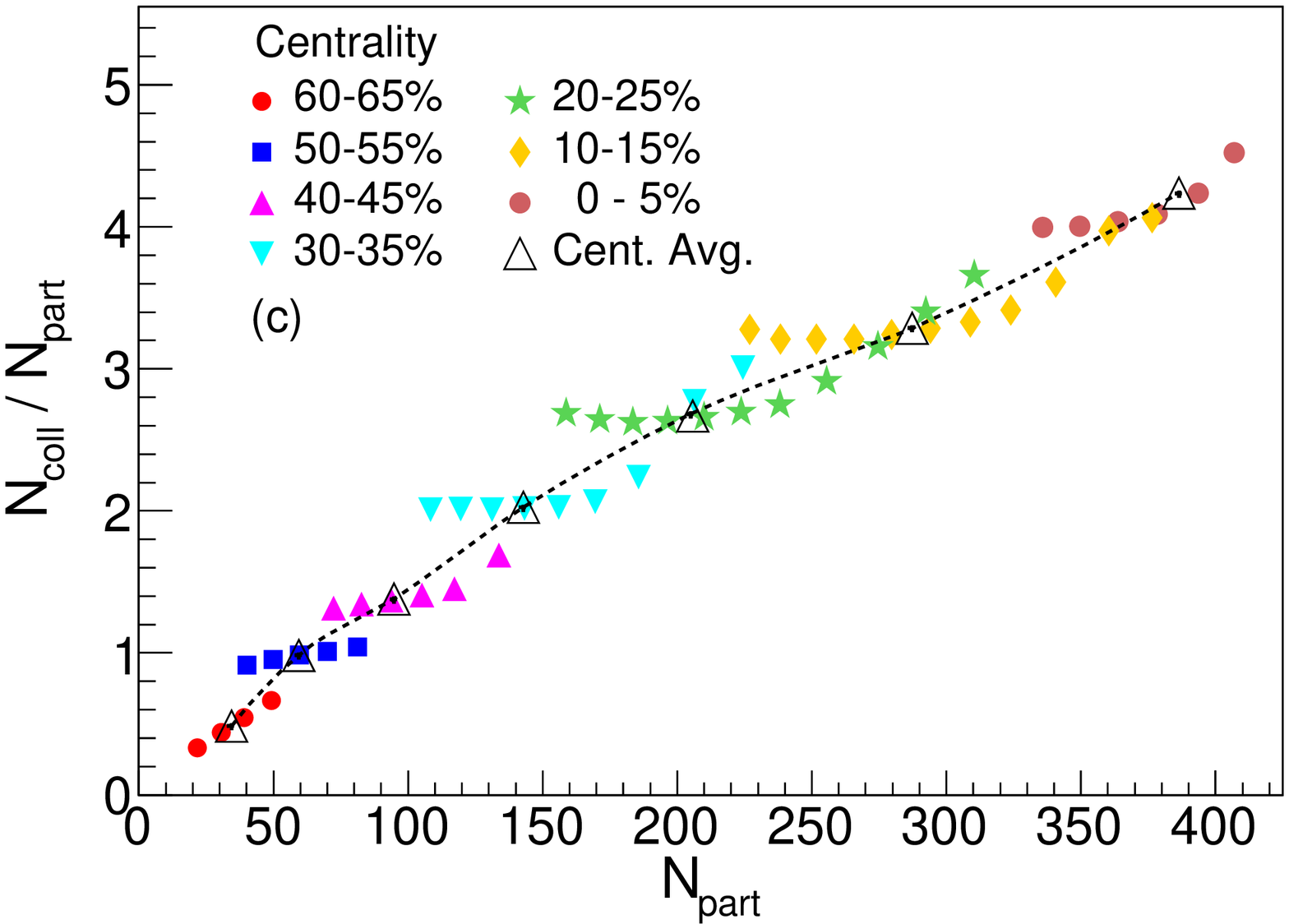} 
  \includegraphics[width=0.45\textwidth]{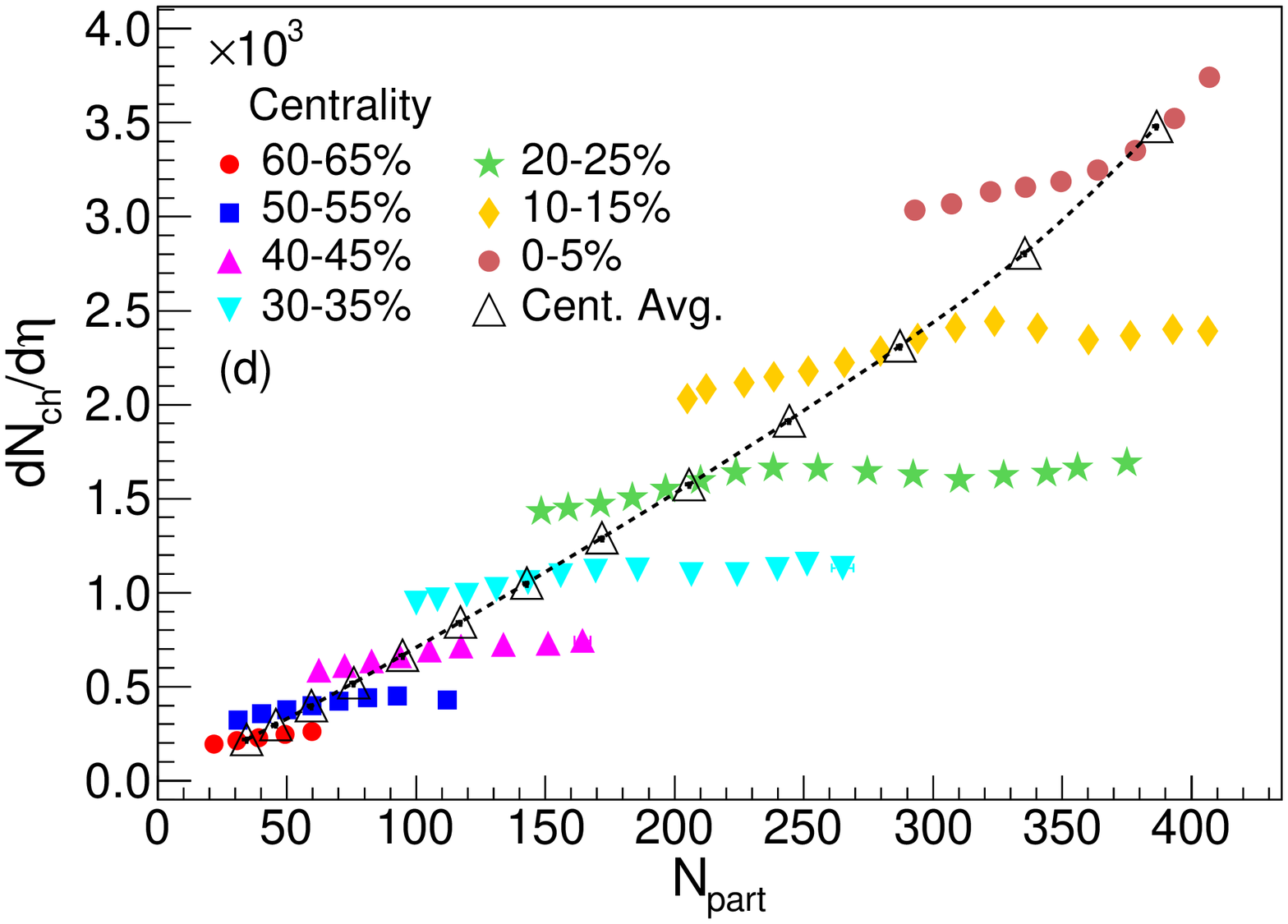}
  }    
\caption{(Color online) Different qualitative dependence of $\sigma_x$, $\sigma_y$, $N_{\text{coll}}$ 
and $dN_{ch}/d\eta$ on $N_{\text{part}}$ with centrality bins and $L+R$ bins.}
\label{fig.IS}
\end{center}
\end{figure*}

We will now present the results of our analysis based on the spectator binning on top of the multiplicity binning.
We show the results from centrality binning alone by the open traingles joined by a dotted line. Here we propose 
to further bin each such centrality bin into different $L+R$ spectator bins as well. These are shown by the colored 
symbols, each color representing a centrality bin. 

Due to E/E fluctuation of the participant positions, the principal axes of inertia of the participant (P) 
nucleons is shifted as well as tilted with respect to those of the nucleus-nucleus (N) system~\cite{Bhalerao:2006tp}. 
Hence, we first perform the necessary translation, 
\beqa
x' &=& x_N - \la x_N\ra,\label{eq.translationx}\\
y' &=& y_N - \la y_N\ra,\label{eq.translationy}
\eeqa
where $\l x_N,\text{ }y_N\r$ denote the nucleon coordinates in the N coordinate system, and further rotate the 
primed coordinate system by the second order participant plane angle $\Psi_2^{\text{PP}}$, so as to coincide with the 
P coordinate system. $\Psi_2^{\text{PP}}$ is obtained as follows,
\beqa
\varepsilon_ne^{i\Psi_n^{\text{PP}}} &=& \frac{\sum_i {r'}_i^ne^{i n\phi'_i}}{\sum_i {r'}_i^n},\label{eq.rotation}
\eeqa
where $\l r'_i,\phi'_i\r$ is the new 2-D polar coordinate of the $i$th participant in the primed coordinate 
system. $\Psi_2^{\text{PP}}$ is obtained for $n=2$. The initial spatial geometry of the overlap region is encoded 
in the eccentricities $\varepsilon_n$ defined in Eq.~\ref{eq.rotation} of which $\varepsilon_2$ and $\varepsilon_3$ will 
be discussed in this work. 

We will first analyse the distribution of the participants in the plane transverse to the collision axis, 
which we call the collision plane. In Figs.~\ref{fig.IS} (a) and \ref{fig.IS} (b) we have 
plotted the bin average of the standard deviation in the $x_P$ ($\sigma_x=\sqrt{\la x_P^2\ra - \la x_P\ra^2}$) and 
$y_P$ ($\sigma_y=\sqrt{\la y_P^2\ra - \la y_P\ra^2}$) coordinates of the participants measured with respect to the 
P coordinate system. $\sigma_x$ and $\sigma_y$ gives us an idea of the initial size of the fireball 
on the collision plane at the time of collision. The curves for $\sigma_x$ and $\sigma_y$ vs $N_{\text{part}}$ show 
different correlations along centrality and $L+R$ bins. They both decrease more rapidly along $L+R$ bins than along 
centrality bins. This brings us to Fig.~\ref{fig.IS} (c) where the number of binary collisions, 
$N_{\text{coll}}$ is shown normalised to $N_{\text{part}}$, the number of participants. Within a geometry based approach of particle 
production and initial conditions in HICs, e.g. the two component Glauber model, $N_{\text{part}}$ and 
$N_{\text{coll}}$ are the two most essential ingredients that determine the IS as well as the FS 
multiplicity~\cite{Bialas:1976ed,Wang:2000bf,Kharzeev:2000ph}. 
A collision between two nucleons (one from each of the colliding nucleus A and B) with coordinates $\l x_A,y_A\r$ and 
$\l x_B,y_B\r$ on the collision plane occurs if they satisfy the following simple geometrical criteria
\beq
\l x_A-x_B\r^2+\l y_A-y_B\r^2\leq \frac{\sigma_{NN}}{\pi}\label{eq.collide}
\eeq
where $\sigma_{NN}$ is the nucleon-nucleon cross section. $N_{\text{part}}$ is the sum of all the nucleons that 
satisfy Eq.~\ref{eq.collide}. $N_{\text{coll}}$, on the other hand, is the sum of all such possible binary collisions 
between the participants. The distribution of the participants on the collision plane determine the 
IS eccentricities given by Eq.~\ref{eq.rotation}.  
As shown in Fig.~\ref{fig.IS} (c), the $L+R$ bins of a higher centrality bin mostly have larger value of $N_{\text{coll}}$ 
compared to a lower centrality bin. This is so because as seen in Figs.~\ref{fig.IS} (a) and (b), bins with same 
$N_{\text{part}}$ but higher centrality occupy a smaller area on the collision plane as implied by smaller values of 
$\sigma_x$ and $\sigma_y$. This means the participants are more lined up along the beam axis than perpendicular to it, hence having 
a higher value for $N_{\text{coll}}$ and smaller $\sigma_x$ and $\sigma_y$. Finally in Fig.~\ref{fig.IS} (d) we have 
plotted $dN_{\text{ch}}/d\eta$ vs $N_{\text{part}}$. Clearly, the correlations along centrality bins is different 
from that along the $L+R$ bins. In a given centrality bin as $L+R$ increases, $N_{\text{part}}$ decreases. 
However, $dN_{ch}/d\eta$ almost remains constant which in turn implies constancy of the initial energy deposited even 
though $N_{\text{part}}$ decreases. This suggests that in a given centrality, as we go towards bins with larger $L+R$, 
the energy deposit pattern changes- one expects to find larger energy gradients and more number of energy hot spots (as the same 
energy is deposited over a smaller transverse area by lesser $N_{\text{part}}$). This should result in very different 
viscous effects as one scans over bins with varying $L+R$ at a given centrality which will be best borne out by 
observables on anisotropic flow.

\begin{figure*}[htb]
 \begin{center}
  \scalebox{1}{
  \includegraphics[width=0.4\textwidth]{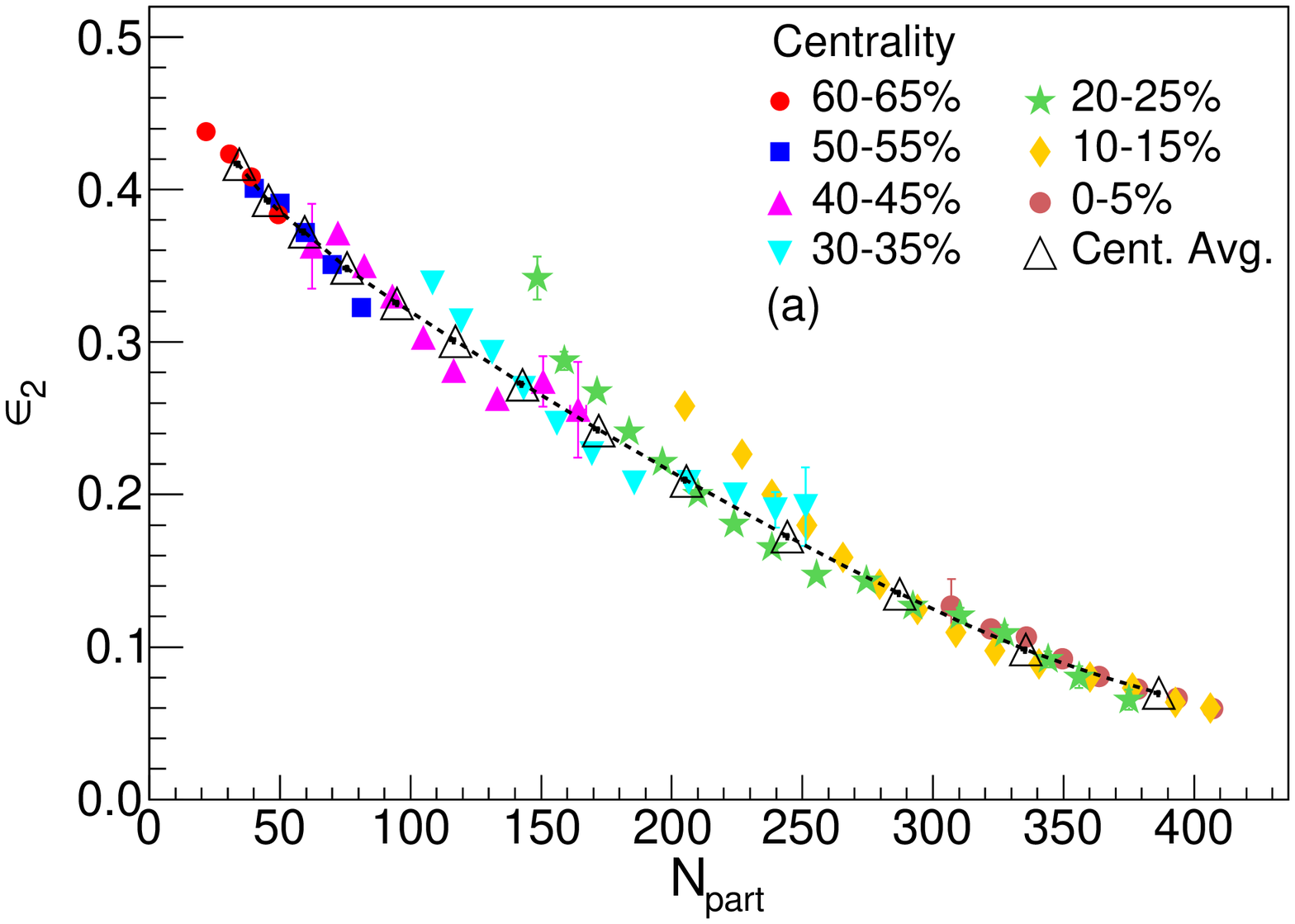}
  \includegraphics[width=0.4\textwidth]{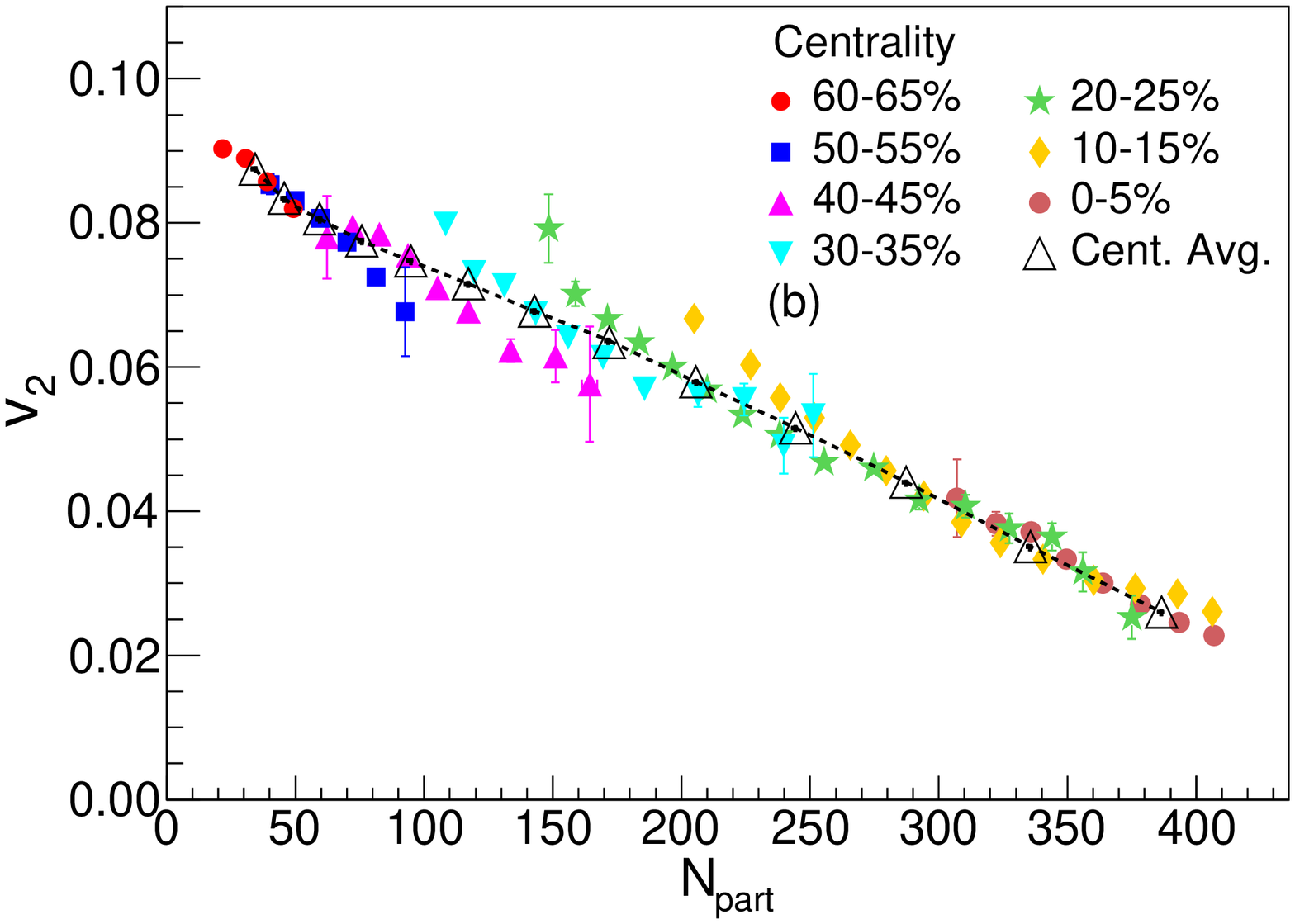}
  }\\
  \scalebox{1}{ 
  \includegraphics[width=0.4\textwidth]{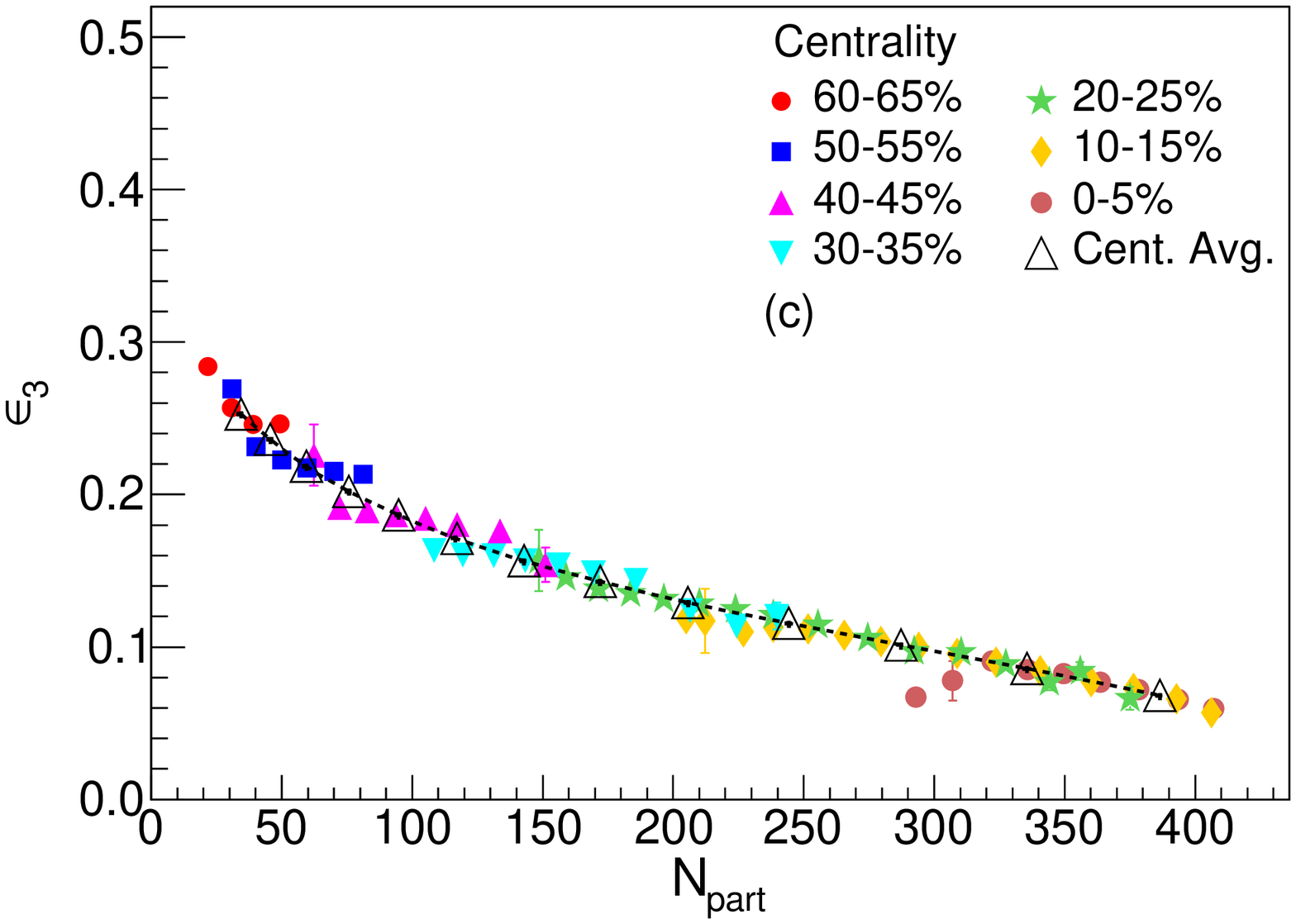}  
  \includegraphics[width=0.4\textwidth]{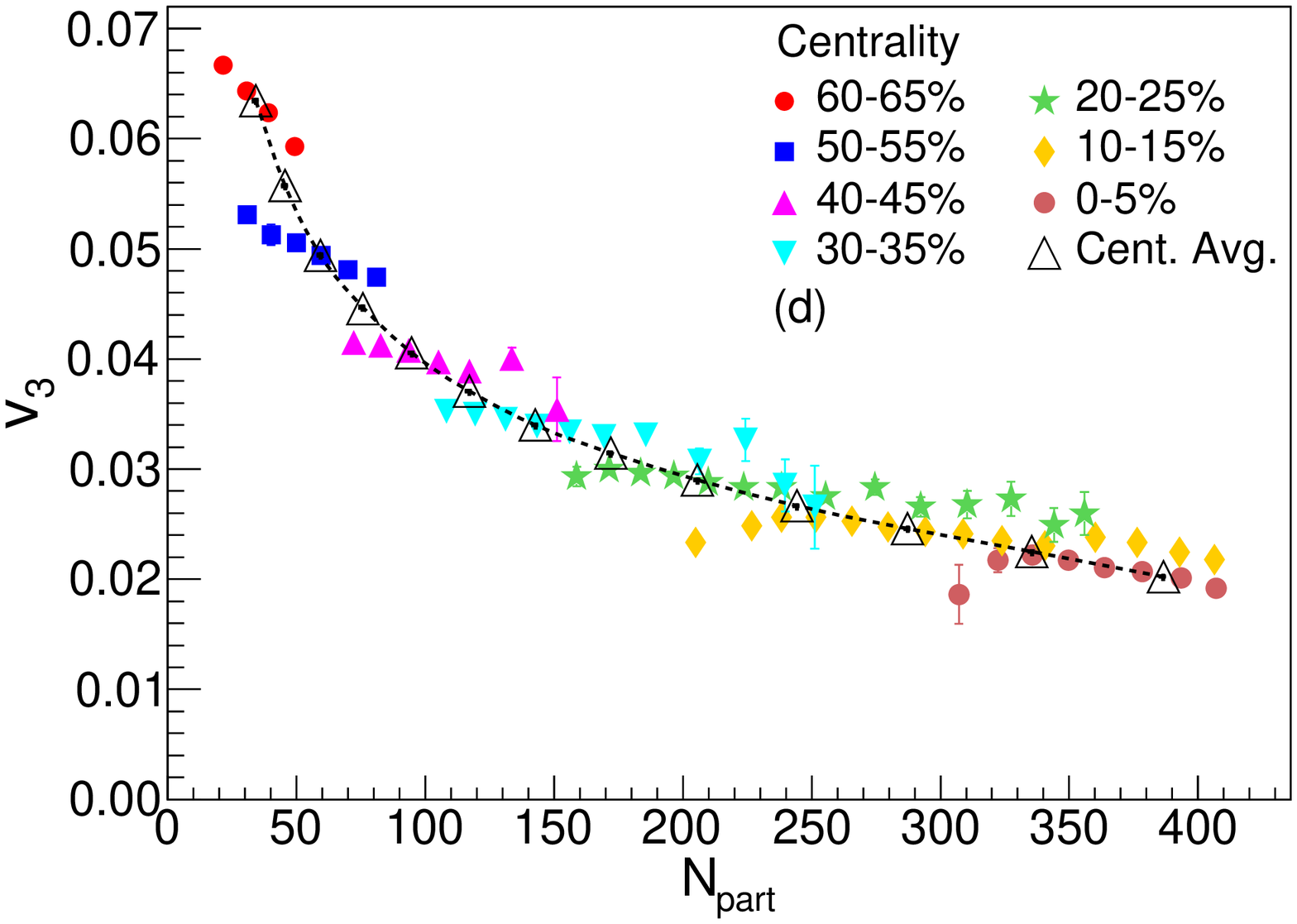} 
  }
   \scalebox{1}{ 
   \includegraphics[width=0.4\textwidth]{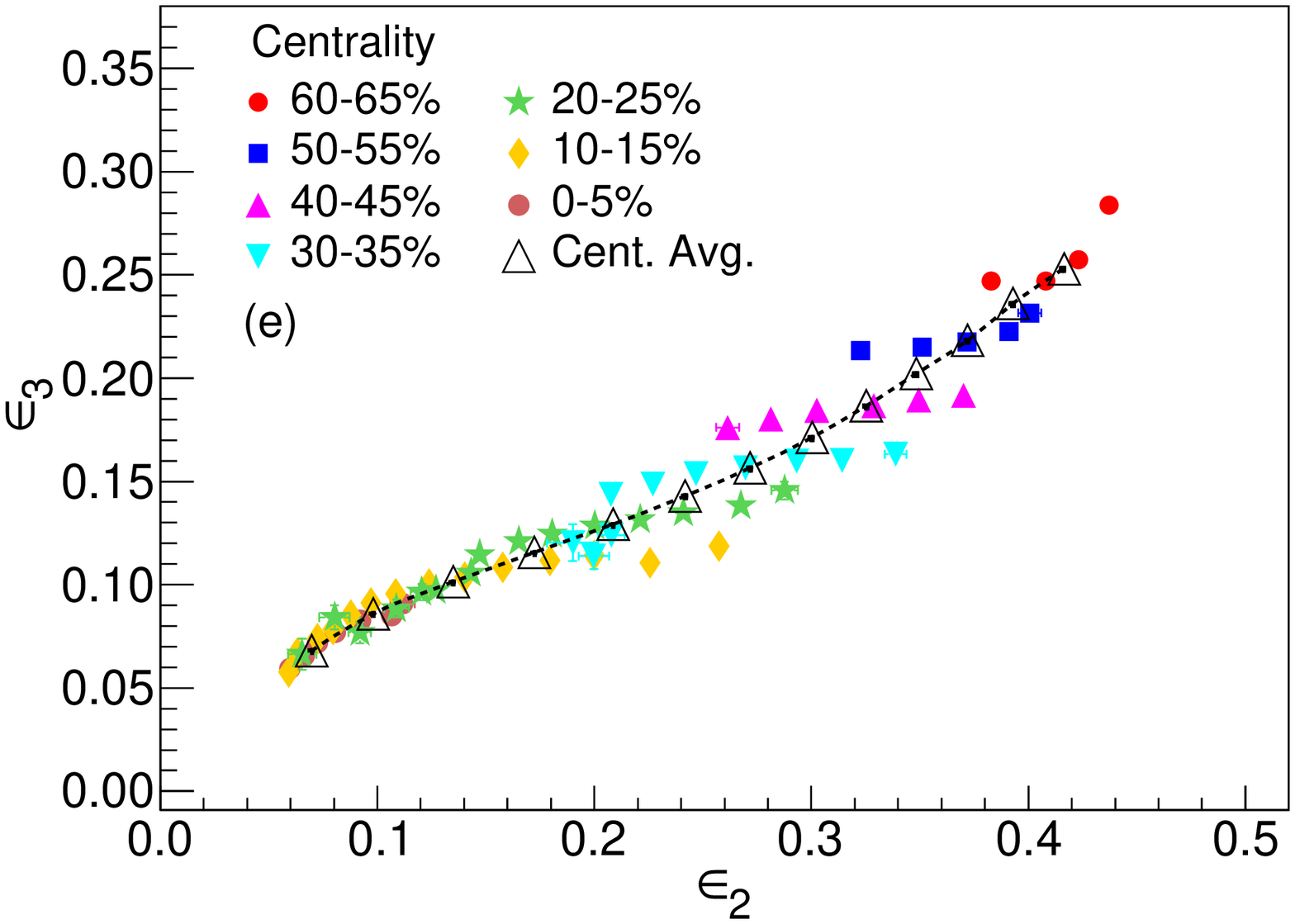}
   \includegraphics[width=0.4\textwidth]{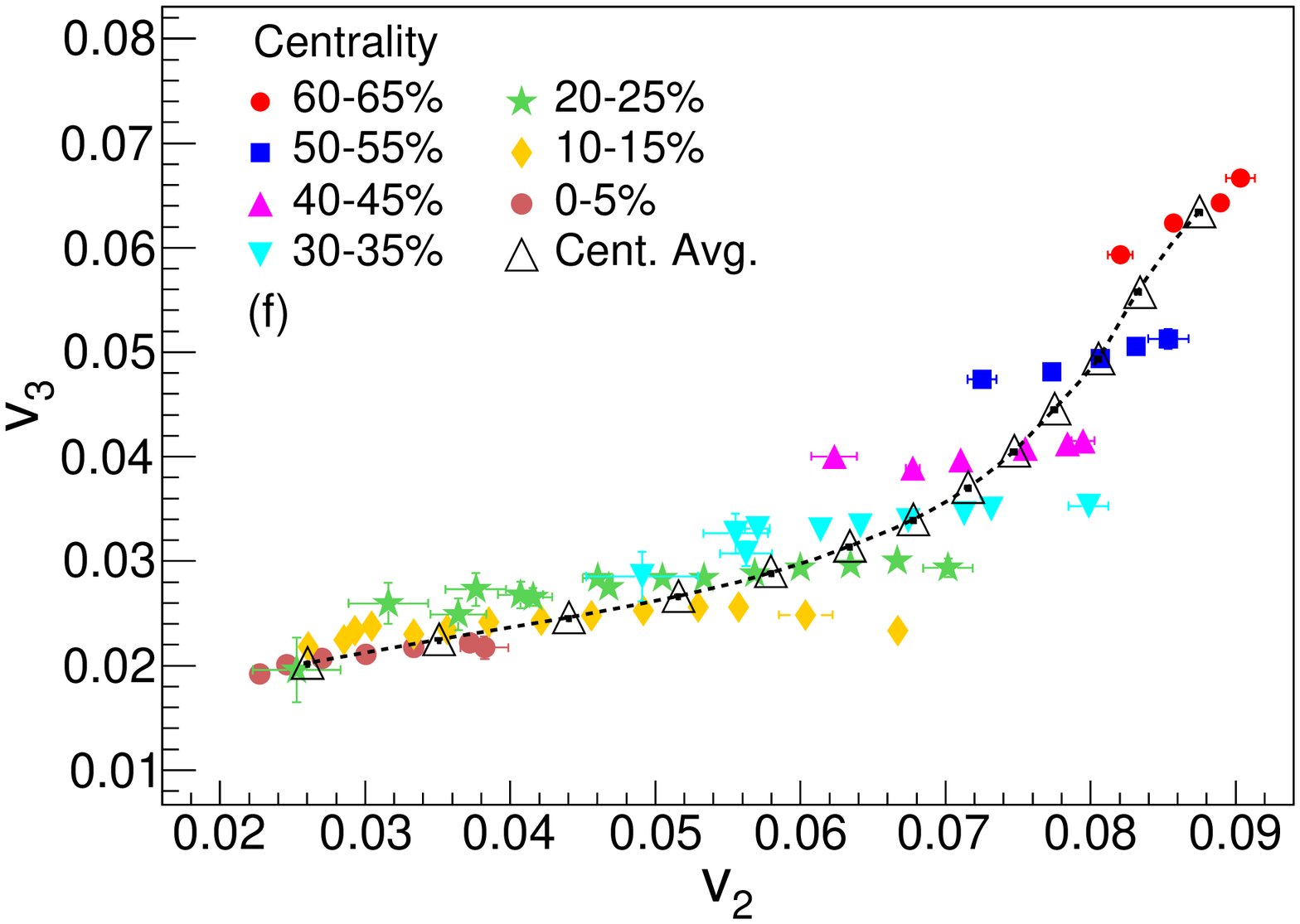}
   }
\caption{(Color online) Different qualitative dependence of $\varepsilon_2$, $\varepsilon_3$, $v_2$ and 
$v_3$ on $N_{\text{part}}$ with centrality bins and $L+R$ bins are shown. We also show the correlation 
between $\varepsilon_2$ and $\varepsilon_3$ as well as between $v_2$ and $v_3$ with centrality and $L+R$ bins.}
\label{fig.Geom}
\end{center}
\end{figure*}

The collective hydrodynamic response converts the IS spatial anisotropy of the fireball 
as reflected by $\varepsilon_n$ into the FS azimuthal anisotropy of the produced charged particle characterised 
by the flow observables $\l v_n,\Psi_n^{EP}\r$,
\beqa
\frac{dN}{d\phi} &\propto& 1+2\sum_{n=1}^{\infty}v_n\cos\l n\l\phi-\Psi_n^{EP}\r\r
= 1+2\sum_{n=1}^{\infty}\l v_{n,x}\cos\l n\phi\r+ v_{n,y}\sin\l n\phi\r\r,\label{eq.flow1}\\
Q_{n,x}&=&\sum_i^M\cos\l n\phi_i\r,\quad Q_{n,y}=\sum_i^M\sin\l n\phi_i\r,\label{eq.Qn}\\
v_{n,x}&=&\frac{1}{M}Q_{n,x},\qquad\qquad v_{n,y}=\frac{1}{M}Q_{n,y},\qquad\qquad v_n=\sqrt{v_{n,x}^2+v_{n,y}^2}\label{eq.flow2}
\eeqa
Here $\phi_i$ is the azimuthal angle of the $i$th particle, $M$ is the total number of particles and $\Psi_n^{EP}$ 
is the event plane angle, measured using the produced particles~\cite{Poskanzer:1998yz}. There have been attempts to 
study the influence of the initial event shape on the ensuing fireball dynamics and hence 
on various FS observables~\cite{Schukraft:2012ah, Huo:2013qma}. We now focus our attention on the event shape- how $L+R$ binning 
introduces a control parameter to tune the IS geometry.

 \begin{figure*}[htb]
  \begin{center}
   \scalebox{1}{
    \includegraphics[width=0.45\textwidth]{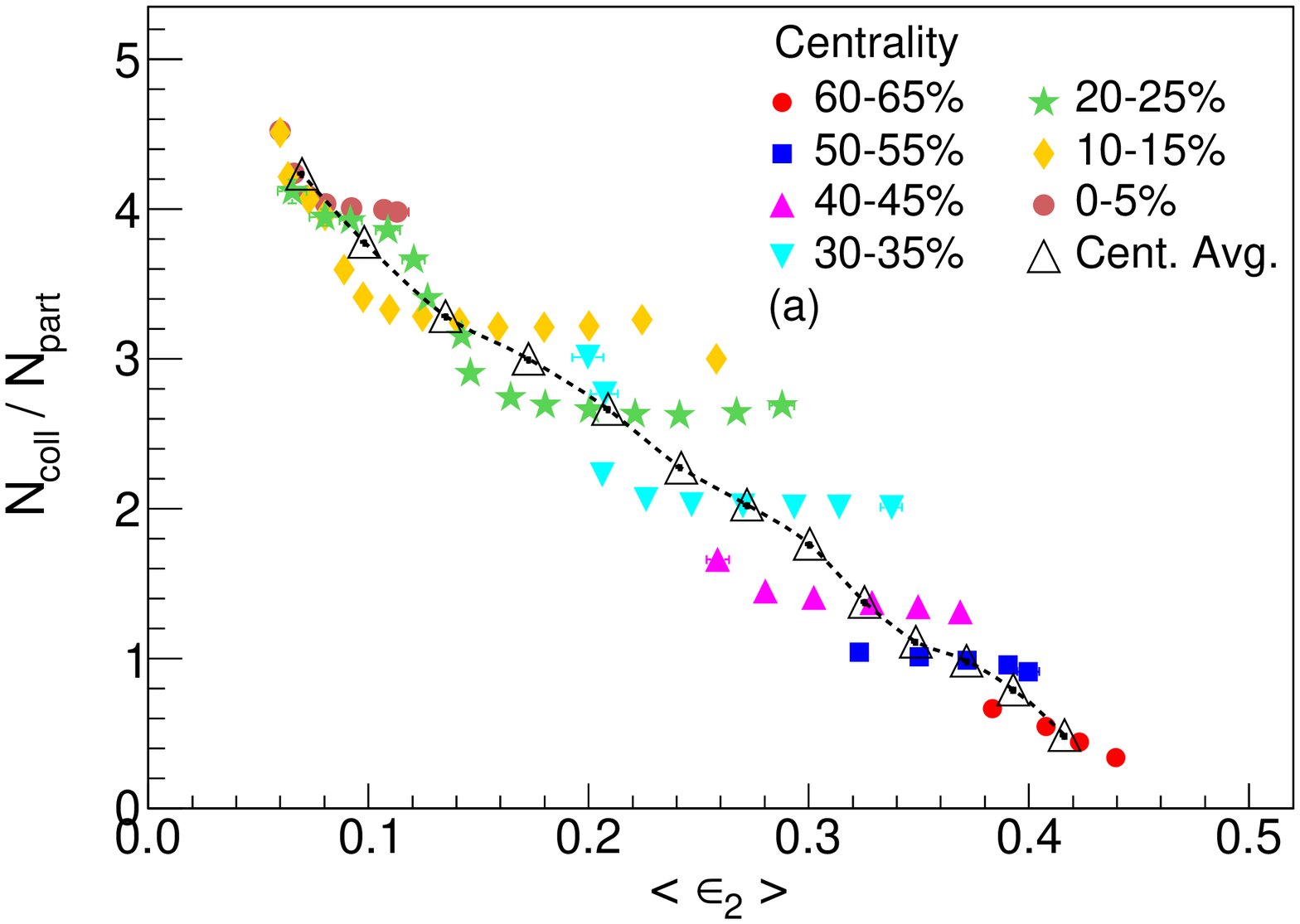}
    \includegraphics[width=0.45\textwidth]{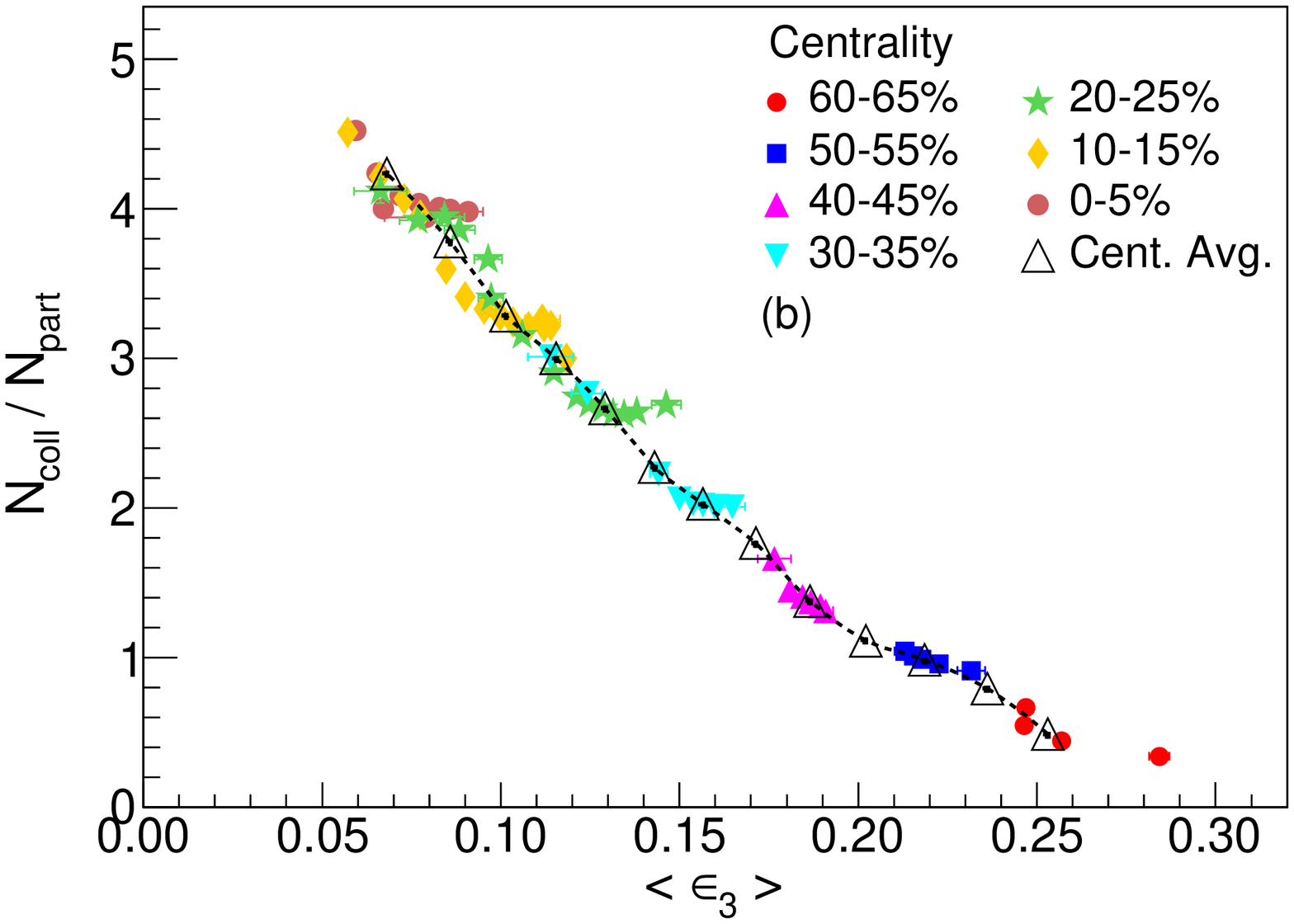}
   }
 \caption{(Color online) The correlation between $\varepsilon_2-N_{coll}/N_{part}$ and $\varepsilon_3-N_{coll}/N_{part}$ 
 for different centrality and $L+R$ bins.}
 \label{fig.ncollgeom}
 \end{center}
 \end{figure*} 

From Figs.~\ref{fig.Geom} (a) and (c) it is clear that $\varepsilon_2$ and $\varepsilon_3$ can 
be tuned by simply triggering on different $L+R$ bins within a particular centrality bin. Thus the spectator 
bins allow a direct access to the initial event shape. $\varepsilon_2$ and $\varepsilon_3$ show almost similar 
variation along centrality and spectator bins for central events. For mid-central to peripheral bins, starting 
from $\l20-25\r\%$ centrality bin, the correlation between $\varepsilon_2$ and $N_{\text{part}}$ 
along spectator bins is slightly steeper compared to that along centrality bins. On the other hand, $\varepsilon_3$ correlation 
with $N_{\text{part}}$ is gentler along $L+R$ bins. This difference in variation of $\varepsilon_2$ and $\varepsilon_3$ along 
$L+R$ vs centrality bins becomes more transparent in the FS flow observables, $v_2$ and $v_3$ as shown in 
Figs.~\ref{fig.Geom} (b) and (d). We note that the $L+R$ bins preserve the ususal linear relation 
between $\l\varepsilon_2,\varepsilon_3\r$ and $\l v_2,v_3\r$. This also gives rise to the $\varepsilon_2-\varepsilon_3$ 
and $v_2-v_3$ correlation plots as shown in Figs.~\ref{fig.Geom} (e) and (f). 
It is clear that the spectator binning allows us to access novel geometries in terms of 
$\l\varepsilon_2,\varepsilon_3\r$ pairs which are never accessible in centrality binning 
alone. Moreover, since the trends are quite different along spectator bins, they cannot be accessed even if one performs 
a narrower centrality binning. For $\l20-25\r\%$ and more peripheral bins, there is much smaller correlation 
along $L+R$ bins as compared to that along centrality bins, e.g. in case of the $\l20-25\r\%$ centrality bin while 
$v_3$ changes by only $10\%$ for different $L+R$ bins, $v_2$ varies by around $130\%$. This will allow us to disentangle the 
effects of $v_2$ from $v_3$ on other observables, e.g. the non-linear mode couplings during the hydrodynamic expansion
that result in correlations in $\l v_2 - v_4\r$ and $\l v_2,v_3 - v_5\r$ could be studied. In this regard, it 
will be interesting to look at correlation plots of $\l v_2,v_4\r$, $\l v_2,v_5\r$ and $\l v_3, v_5\r$ in data with 
a combined binning in $dN_{ch}/d\eta$ and $L+R$ bins. Thus, a combined binning in terms of centrality and 
$L+R$ allows us to disentangle the contribution of different IS characteristics on the FS observables. Here we have discussed 
two such cases, (a) $N_{\text{part}}$ and $N_{\text{coll}}$ and their contribution towards FS $dN_{ch}/d\eta$ and 
(b) $\varepsilon_2$ and $\varepsilon_3$ and their contribution towards $v_2$ and $v_3$ respectively. In 
Fig.~\ref{fig.ncollgeom} we show the correlation between $N_{coll}/N_{part}$ with $\varepsilon_2$ and $\varepsilon_3$. 
Thus with the introduction of $L+R$ bins, we can now study the evolution of similar initial 
geometry ($\varepsilon_2$, $\varepsilon_3$) but with different mechanism of energy deposition ($N_{coll}/N_{part}$).

Thus studies with $L+R$ bins can complement ongoing studies with $q_2$ bins which also aim at studying the influence of the 
event shape on various FS observables~\cite{Schukraft:2012ah, Huo:2013qma} where $q_2$ is obtained from the the 
second harmonic flow vector $Q_2$~\cite{Poskanzer:1998yz,Schukraft:2012ah}
\beqa
Q_2=\sqrt{Q_{2,x}^2+Q_{2,y}^2},\quad q_2=\frac{1}{\sqrt{M}}Q_2\label{eq.q2}
\eeqa
Here $Q_{2,x}$ and $Q_{2,y}$ are computed using particles produced in the FS. It is important to note that unlike 
the $q_2$ binning procedure, in this method of studying the IS 
geometry through $L+R$ bins the linearity relationship between $\l v_2,v_3\r$ and $\l\varepsilon_2,\varepsilon_3\r$ 
is not essential as the spectators, being IS observables, provide a direct access to the IS geometry.

 \begin{figure*}[htb]
  \begin{center}
   \scalebox{1}{
    \includegraphics[width=0.45\textwidth]{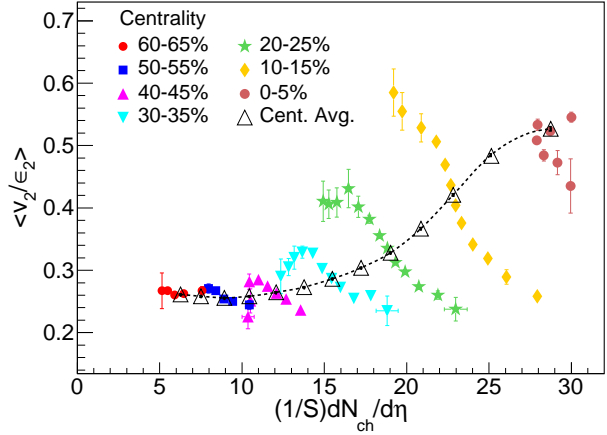}
   }
 \caption{(Color online) $v_2/\varepsilon_2$ vs $\l1/S\r dN_{ch}/d\eta$ for different centrality and $L+R$ bins. The $L+R$ bins break the 
 scaling relation between $v_2/\varepsilon_2$ and $\l1/S\r dN_{ch}/d\eta$ that is exhibited by centrality bins.}
 \label{fig.v2bye2}
 \end{center}
 \end{figure*} 
 
 \begin{figure*}[htb]
  \begin{center}
   \scalebox{1}{
    \includegraphics[width=0.45\textwidth]{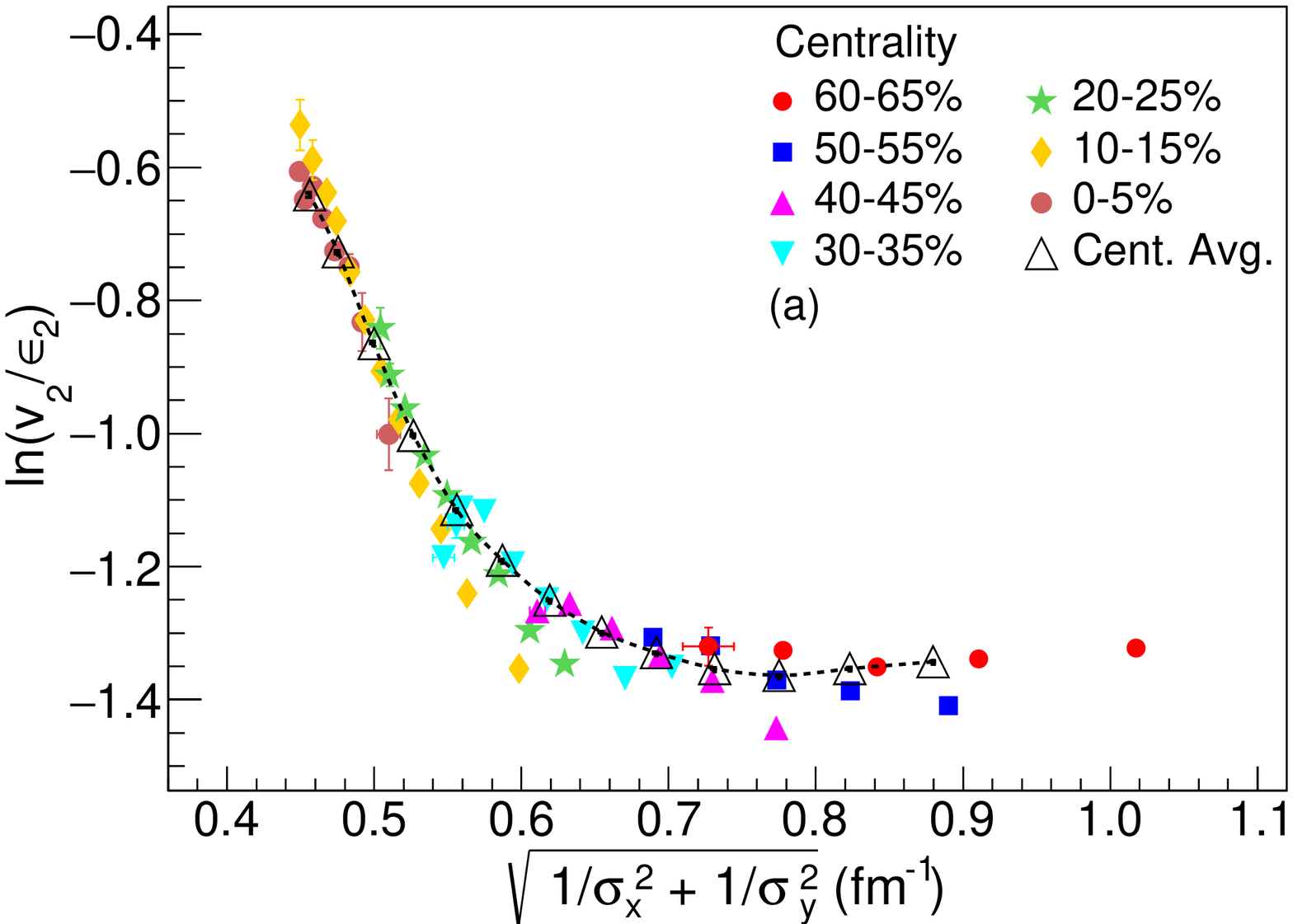}
    \includegraphics[width=0.45\textwidth]{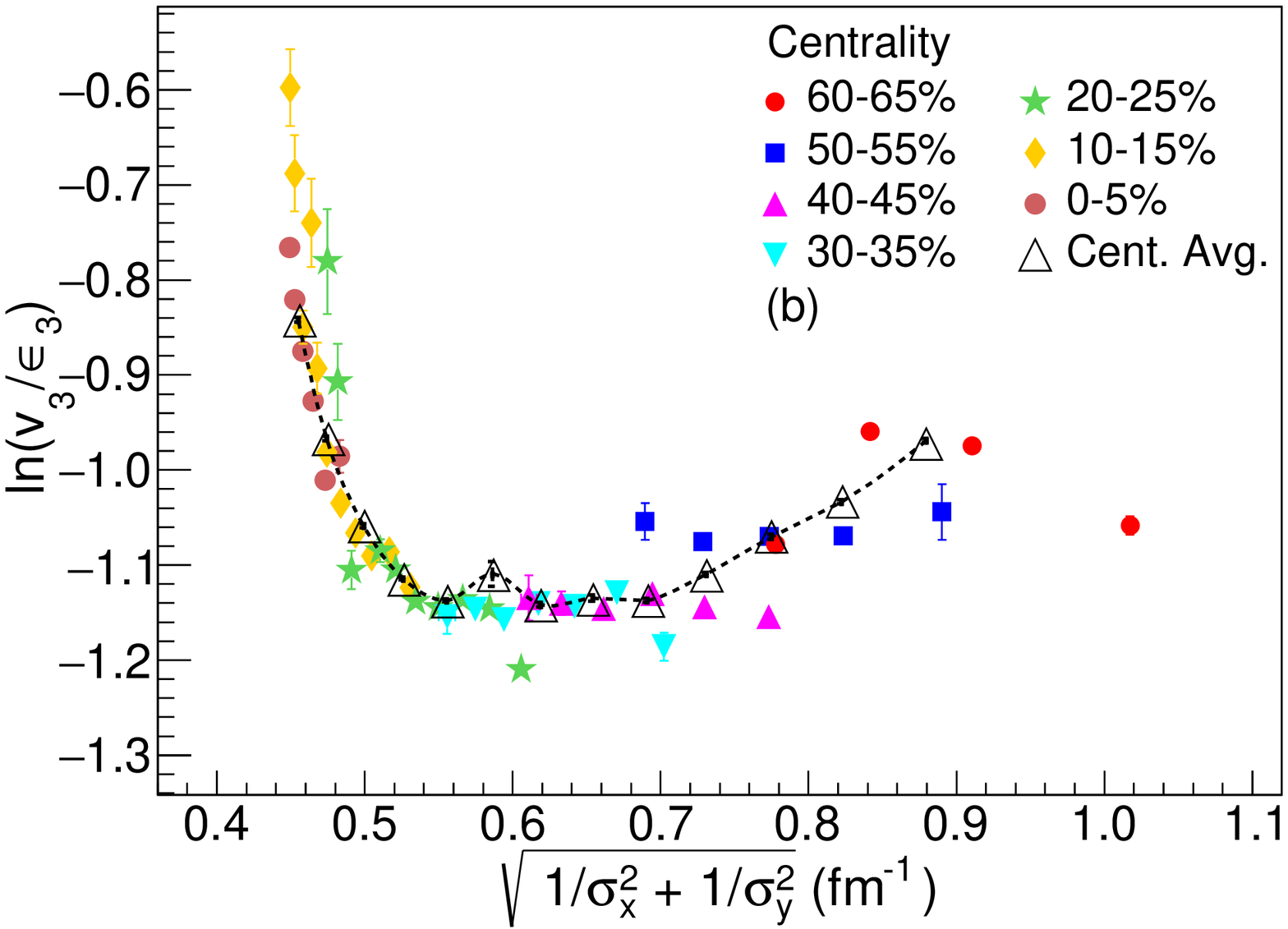}
   }
 \caption{(Color online) Acoustic scaling of the hydrodynamic response $\ln\l v_n/\varepsilon_n\r$ vs $1/\Lambda_T$ 
 with $n=2$ (left) and $n=3$ (right) for  different centrality and $L+R$ bins.}
 \label{fig.vnbyen}
 \end{center}
 \end{figure*}  

Ideal fluid hydrodynamics is scale invariant and as a result the final value of the ratio $\l v_2/\varepsilon_2\r$ 
attained is independent of the system size~\cite{Bhalerao:2005mm}. Viscous corrections in a non-ideal fluid 
arising due to incomplete thermalization introduce system size depedence and tend to reduce this ratio. The 
ratio of the microscopic mean free path $\lambda$ to the typical macroscopic system size $\Lambda$, $\lambda/\Lambda$ is 
called the Knudsen number $K$. $K^{-1}$ is expected to be a good measure of the thermalization 
achieved~\cite{Bhalerao:2005mm,Drescher:2007cd}. Since $v_2$ develops over some time during which the fireball 
rapidly expands it is not exactly clear what 
should be the value of $\lambda$ and $\Lambda$ to be used to determine the degree of thermalization. In earlier 
studies, it has been suggested that $v_2$ dominantly develops in the early stage of the fireball evolution. Therefore 
$K^{-1}$ estimated at time $\tau\sim \Lambda_T/c_s$ at the onset of transverse expansion of the fireball, should be used 
as a measure of thermalization~\cite{Bhalerao:2005mm,Drescher:2007cd}. Here $\Lambda_T$ is the transversize size of the fireball 
on the collision plane and $c_s$ is the speed of sound. It was further argued that $K^{-1}$ at $\Lambda_T/c_s$ 
approximately scales with  $\l1/S\r dN_{ch}/d\eta$ where $S$ is the initial transverse 
area on the collision plane. Similar scaling relation between $v_2/\varepsilon_2$ and $\l1/S\r dN_{ch}/d\eta$ is also
expected in the low density regime~\cite{Heiselberg:1998es,Voloshin:1999gs,Kolb:2000fha}. 
Previous analysis of experimental data with centrality binning have indeed found very good scaling relation 
between $v_2/\varepsilon_2$ vs $\l1/S\r dN_{ch}/d\eta$ for different systems like Cu+Cu and 
Au+Au~\cite{Drescher:2007cd,Song:2010mg}. The detailed mechanism of energy deposition in the intial stages of a HIC 
event which ultimately leads to particle production is yet to be understood completely. With centrality as 
the only tuning parameter to separate different initial conditions, it is difficult to discriminate between predictions from 
models with different mechanisms of particle production. Now, with the introduction of $L+R$ bins, we are able to pin 
down the initial conditions more precisely. In Fig.~\ref{fig.v2bye2} we have plotted the ratio $v_2/\varepsilon_2$ 
vs $\l1/S\r dN_{ch}/d\eta$ with centrality and $L+R$ bins. The centrality averaged points exhibit the usual 
trend of an inital fast rise and final saturation of the ratio $v_2/\varepsilon_2$ with $\l1/S\r dN_{ch}/d\eta$. 
However, the $L+R$ bins in each centrality show the opposite behaviour and breaks the usual scaling relation. 
In a given centrality bin, with rise in $L+R$, the transverse overlap area $S$ sharply falls while 
$dN_{ch}/d\eta$ is almost constant. Thus $\l1/S\r dN_{ch}/d\eta$ which acts like a proxy for $K^{-1}$ 
increases with $L+R$ although the hydrodynamic response $v_2/\varepsilon_2$ falls sharply. As mentioned earlier, 
we expect more (less) hot spots and gradients in the initial energy profile of events with larger (smaller) 
$L+R$ leading to more (less) viscous correction. This ultimately leads to inefficient conversion of 
the initial $\varepsilon_n$ to the final $v_n$ in events with larger $L+R$ as compared to those with smaller $L+R$.

The ratio $v_n/\varepsilon_n$ is also expected to exhibit acoustic scaling and receives viscous 
corrections that grow exponentially as $n^2$ and $1/\Lambda_T$~\cite{Staig:2010pn,Lacey:2011ug,Lacey:2013is,Lacey:2013qua}
\beqa
\ln\l\frac{v_n}{\varepsilon_n}\r&\propto&-\frac{4}{3}\frac{n^2\eta}{\Lambda_TTs},\label{eq.acoustic}
\eeqa
with the typical initial transverse size of the system $\Lambda_T$ given by $1/\Lambda_T=\sqrt{1/\sigma_x^2+1/\sigma_y^2}$. 
Such acoustic scaling was found in data across a wide range of beam energies and the shear viscosity to entropy density 
ratio, $\eta/s$ was extracted from the slope of the plot, 
$\ln\l\frac{v_2}{\varepsilon_2}\r$ vs $1/\Lambda_T$~\cite{Lacey:2011ug,Lacey:2013is,Lacey:2013qua}. 
These studies were done with centrality binning alone. Here we perform a consistency check of such scaling laws using 
simulated data by including $L+R$ bins in every centrality bin. In Figs.~\ref{fig.vnbyen} (a) and (b) we 
have plotted $\ln\l v_2/\varepsilon_2\r$ and $\ln\l v_3/\varepsilon_3\r$ respectively vs $1/\Lambda_T$. We 
find approximate simulated data collapse for $\l0-40\r\%$ centrality as well as their corresponding $L+R$ bins. On a closer
scrutiniy, it seems that the slope parameter for bins of different $L+R$ but a particular centrality is different from 
those of a different centrality resulting in a mild breaking of the acoustic scaling. Thus, the introduction of the 
$L+R$ bins enable a more refined extraction of $\eta/s$ from data.

\begin{figure*}[htb]
 \begin{center}
  \scalebox{1}{
  \includegraphics[width=0.3\textwidth]{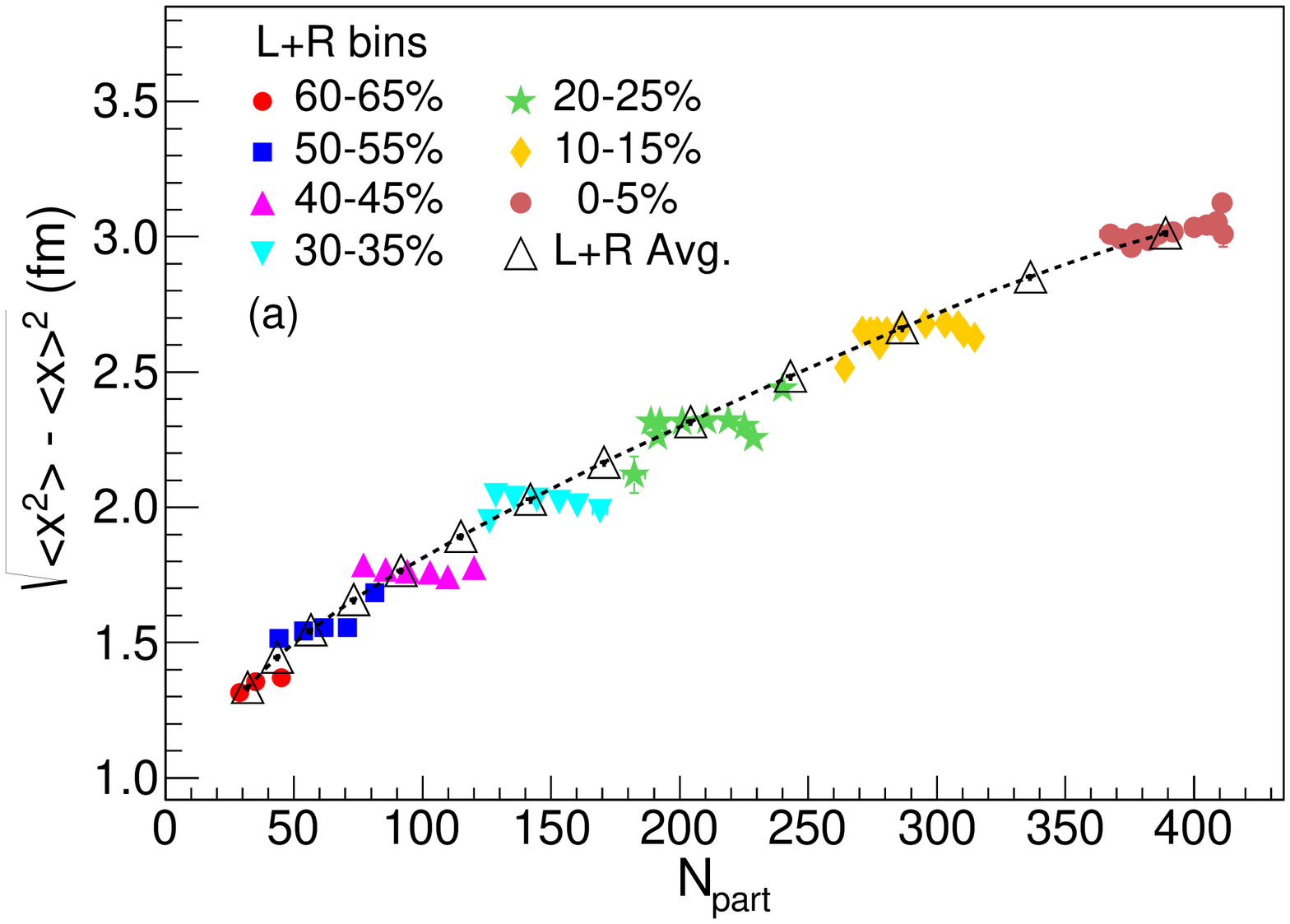}
  \includegraphics[width=0.3\textwidth]{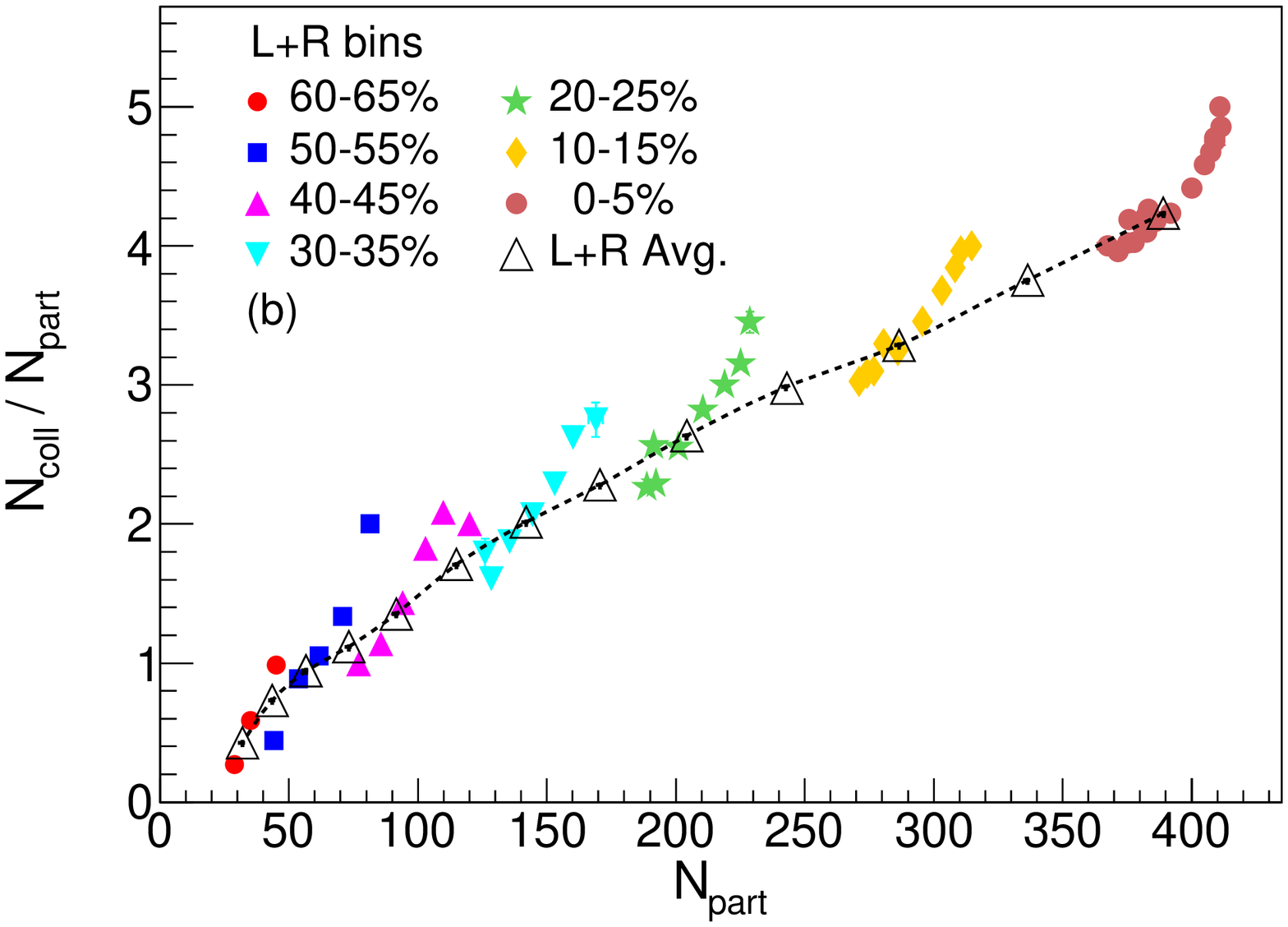}
  \includegraphics[width=0.3\textwidth]{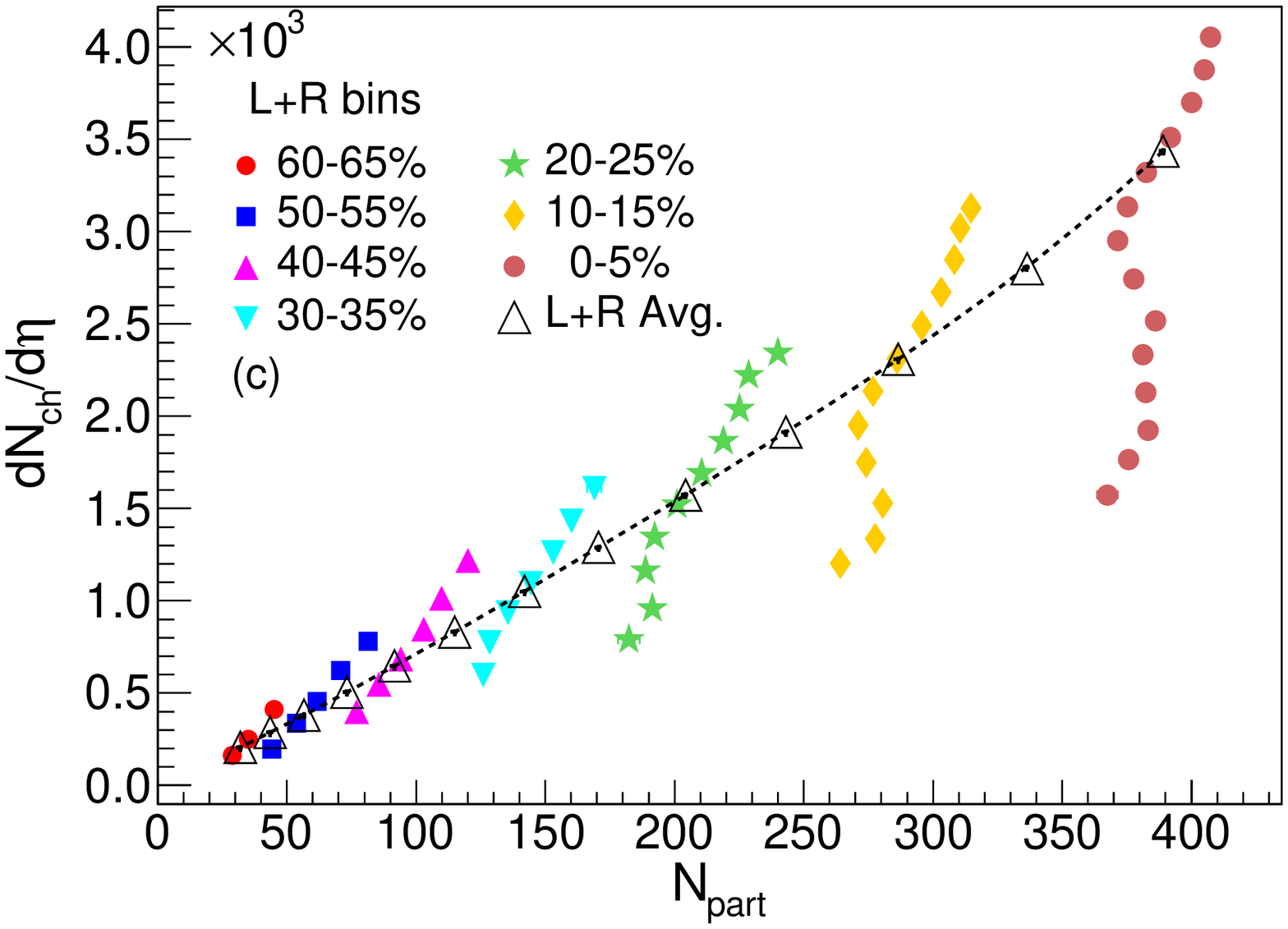}
  }
\caption{(Color online) Different qualitative dependence of $\sigma_x$, $N_{coll}$ and $dN_{ch}/d\eta$ on 
$N_{\text{part}}$ with the reverse binning procedure (first binned by $L+R$ followed by $dN_{ch}/d\eta$) 
are shown.}
\label{fig.reversebin}
\end{center}
\end{figure*}

\begin{figure*}[htb]
 \begin{center}
  \scalebox{1}{
  \includegraphics[width=0.4\textwidth]{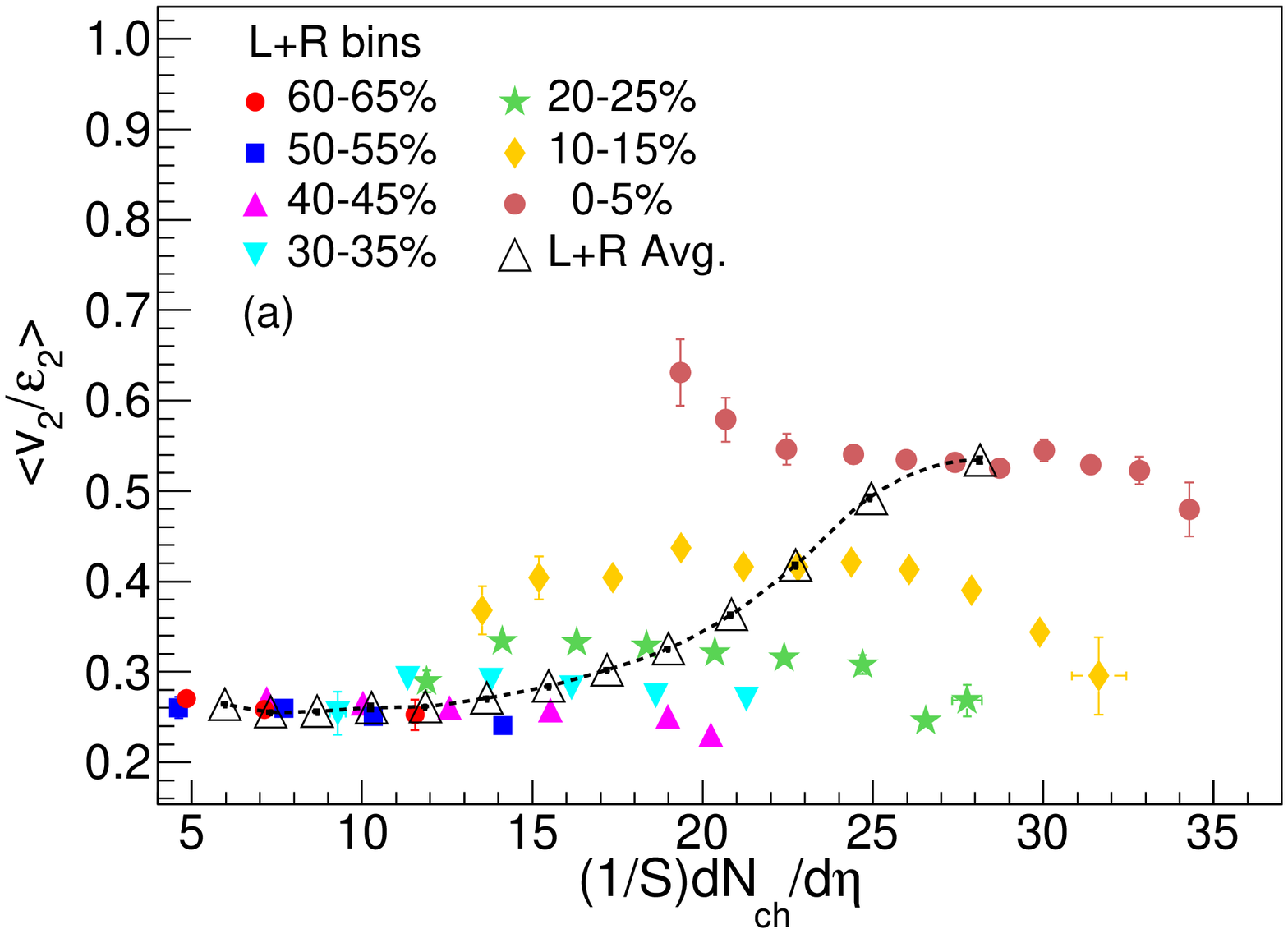}
  \includegraphics[width=0.4\textwidth]{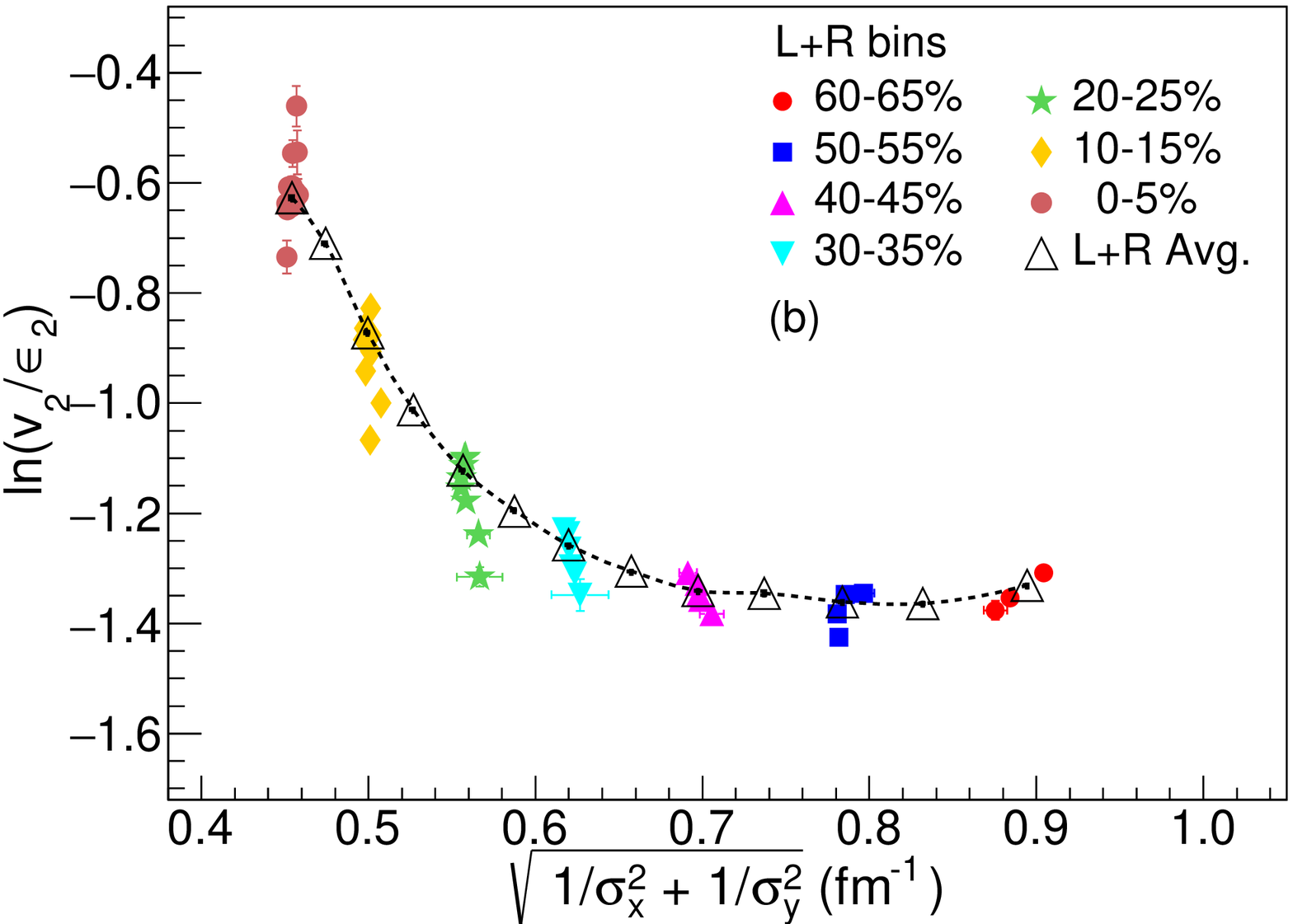}
  }
\caption{(Color online) $v_2/\varepsilon_2$ vs $\l1/S\r dN_{ch}/d\eta$ and $1/\Lambda_T$ with the reverse binning procedure 
(first binned by $L+R$ followed by $dN_{ch}/d\eta$) are shown.}
\label{fig.RBPgeom}
\end{center}
\end{figure*}

So far we have discussed results for events first binned by $dN_{ch}/dy$ followed by $L+R$. If 
$b$ was the only E/E fluctuating quantity, then $b$, $dN_{ch}/dy$ and $L+R$ would have a one to one 
correspondence and hence the final results would be independent of the order of the binning procedure. However, as 
discussed earlier, in HICs there are additional sources of E/E fluctuations apart from the geometrical fluctuation in 
$b$. This means that the final results are sensitive to the order of the binning procedure. In order to illustrate this 
point we have also analysed the events in the reverse binning procedure: first we bin by $L+R$ followed by $dN_{ch}/dy$. 
We have shown a few results in Figs.~\ref{fig.reversebin} and \ref{fig.RBPgeom}. As seen from Fig.~\ref{fig.reversebin} (a), 
$\sigma_x$ which is also a measure of the initial system size is almost constant in a particular $L+R$ bin even though 
$N_{part}$ changes resulting in the strong variation of $N_{coll}/N_{part}$ within a $L+R$ bin as shown in 
Fig.~\ref{fig.reversebin} (b). This finally translates into a stronger variation in $dN_{ch}/d\eta$ within a $L+R$ bin with 
$N_{part}$ as compared to the variation between different $L+R$ bins. This further results in the trends between 
$v_2/\varepsilon_2$ and $1/SdN_{ch}/dy$ as seen in Fig.~\ref{fig.RBPgeom} (a). Within a particular $L+R$ bin, the bins with 
higher $1/SdN_{ch}/dy$ have higher $N_{coll}/N_{part}$ ratio than the average trend resulting in smaller $v_2/\varepsilon_2$ 
compared to the average trend. This trend was seen even in the earlier binning procedure. As seen in Fig.~\ref{fig.RBPgeom} (b), 
the acoustic scaling relation as in Eqn.~\ref{eq.acoustic} does not hold anymore as the slope for the trend of the average 
$L+R$ bins is much softer than the slope along different $dN_{ch}/dy$ bins in the same $L+R$ bin. Thus this breaking of 
the acoustic scaling relation which was mild in the earlier binning procedure becomes much stronger in the reverse 
binning proccedure.

\begin{figure*}[htb]
 \begin{center}
  \scalebox{1}{
  \includegraphics[width=0.4\textwidth]{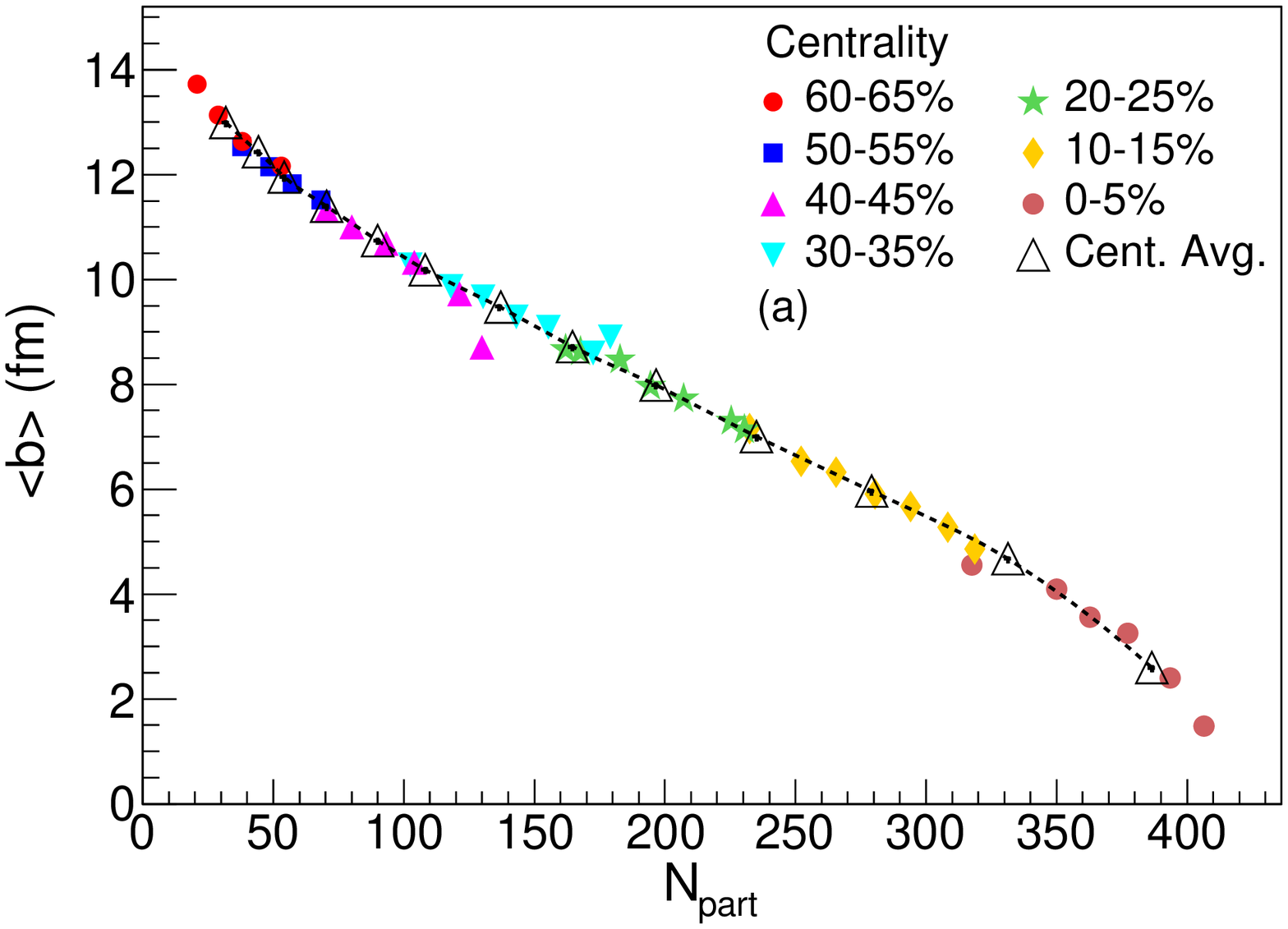}
  \includegraphics[width=0.4\textwidth]{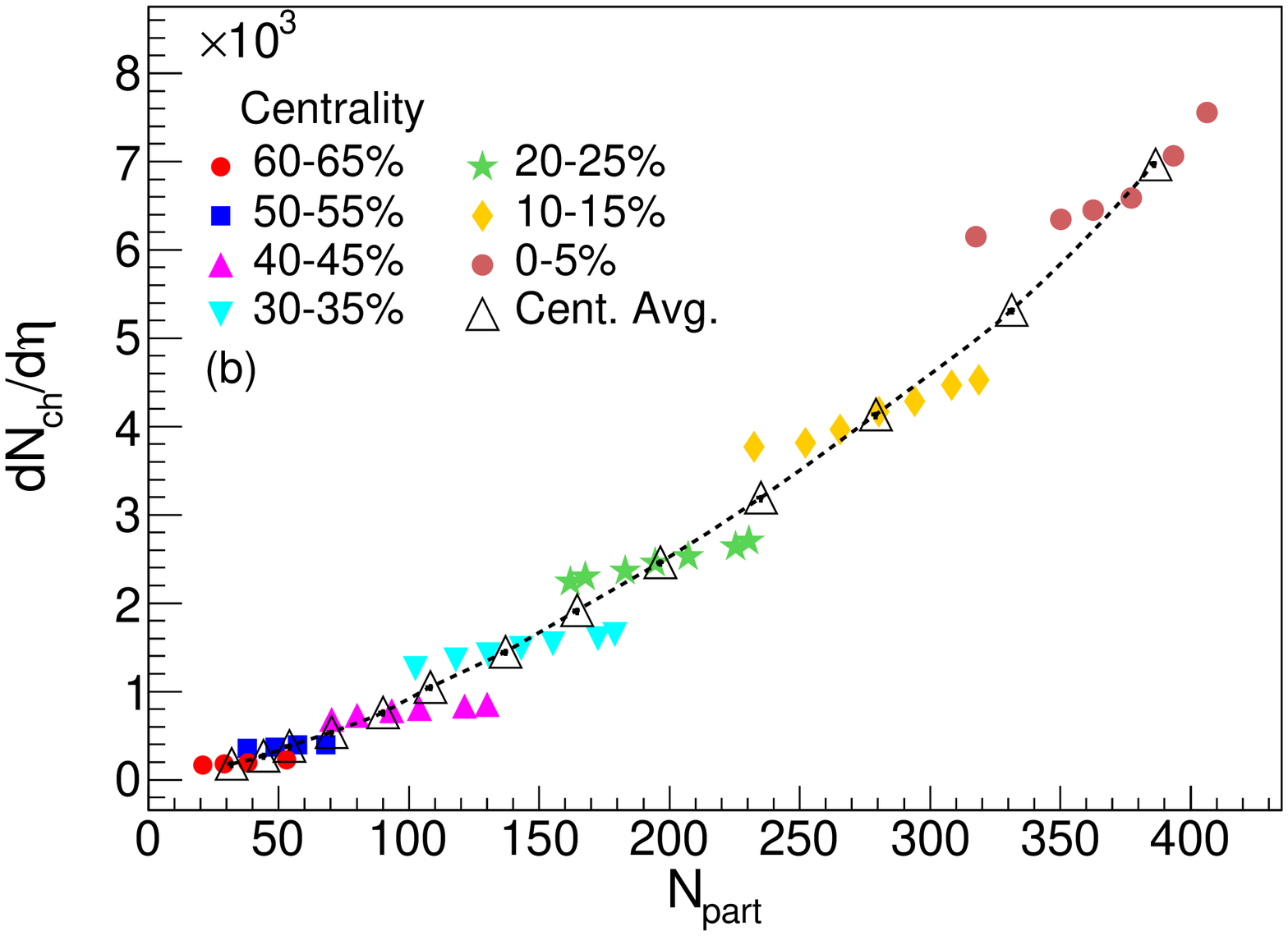}
  }
\caption{(Color online) The trends for $dN_{ch}/d\eta$ and $b$ with $N_{\text{part}}$ as obtained 
in HIJING where the events were binned by $dN_{ch}/d\eta$ followed by $L+R$.}
\label{fig.HIJING}
\end{center}
\end{figure*}

\begin{figure*}[htb]
 \begin{center}
  \scalebox{1}{
  \includegraphics[width=0.4\textwidth]{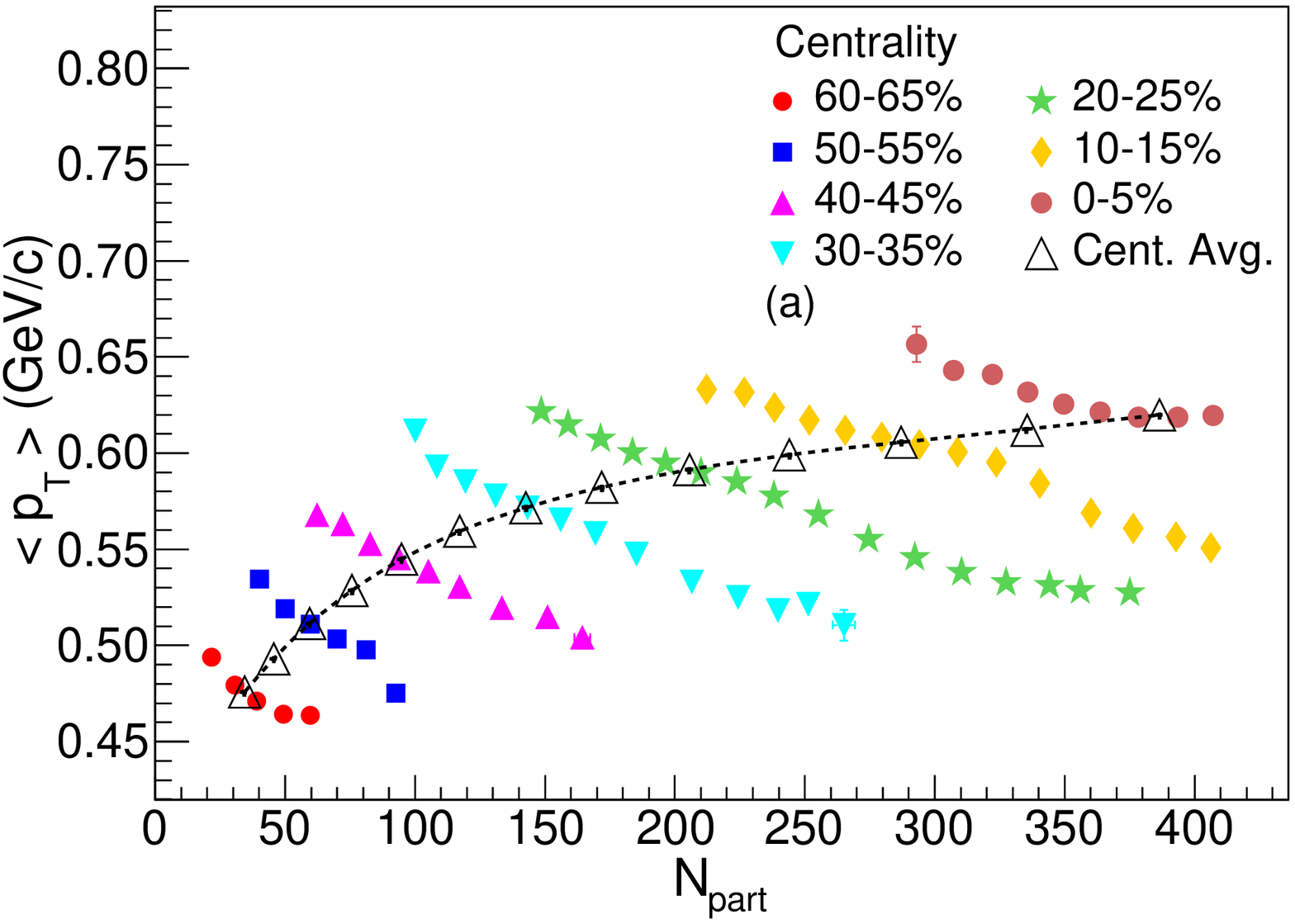}
  \includegraphics[width=0.4\textwidth]{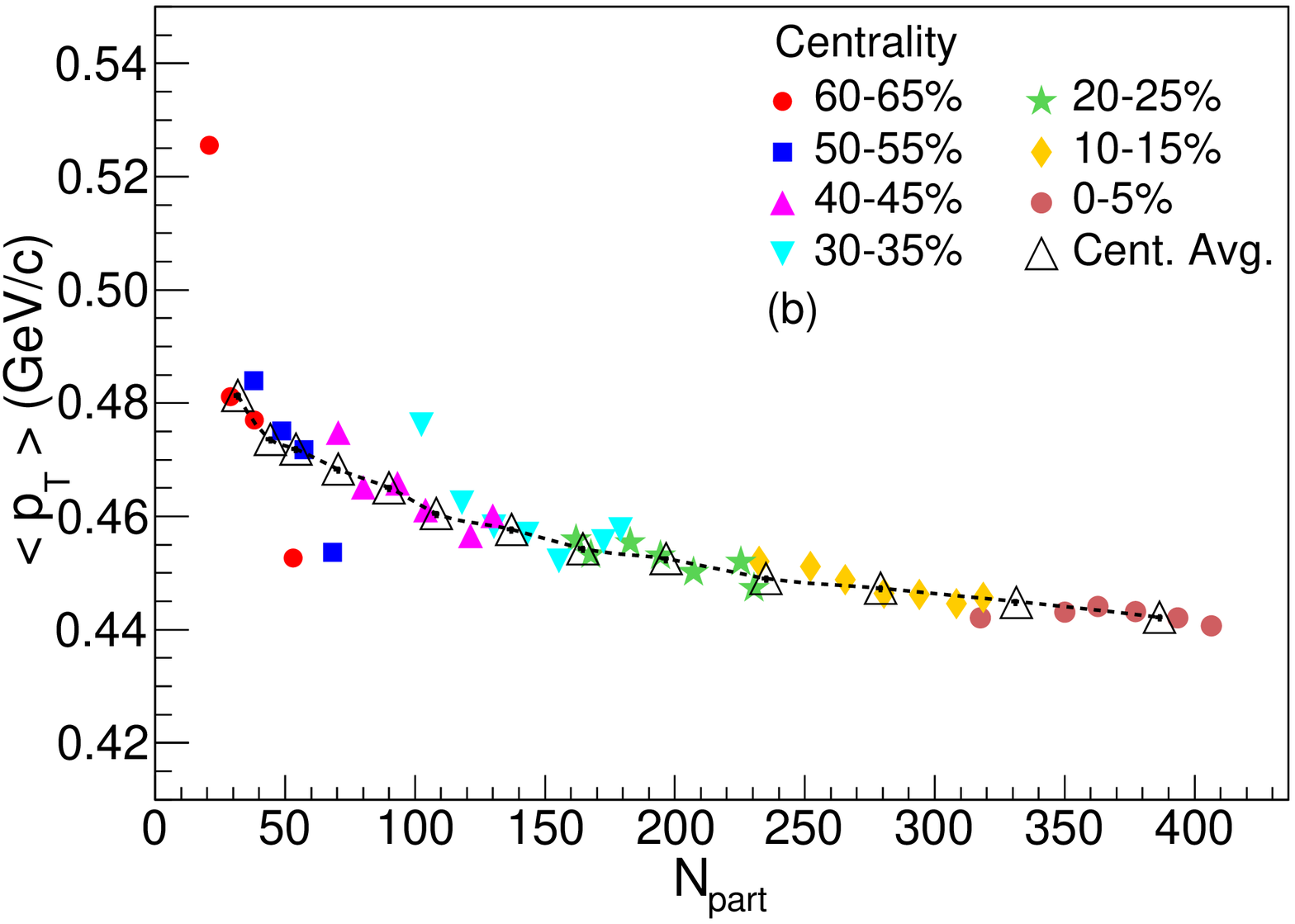}
  }
\caption{(Color online) The trend for $\la p_T\ra$ with $N_{\text{part}}$ as obtained 
in AMPT (left) and HIJING (right) where the events were binned by $dN_{ch}/d\eta$ followed by $L+R$.}
\label{fig.AMPTvsHIJING}
\end{center}
\end{figure*}

So far all our results have been from the AMPT model. We will now show a few results for the combined binning procedure with 
$dN_{ch}/dy$ followed by $L+R$ in HIJING event generator. In Figs.~\ref{fig.HIJING} (a) and (b) we show the results for $b$ 
and $dN_{ch}/dy$ and find them to be very similar to what we obtain in the case of AMPT in Figs.~\ref{fig.zdc} (c) and 
\ref{fig.IS} (d). AMPT takes into account later stage interactions in the partonic as well as hadronic phases while in 
HIJING such interactions are absent. As seen in Fig.~\ref{fig.AMPTvsHIJING}, this results in very different trends for 
$\la p_T\ra$ with $N_{part}$ in AMPT compared to HIJING. In AMPT, we find the $\la p_T\ra$ trends along $L+R$ bins to be 
much different to that of the average trend along different centrality bins with $\sim10-15\%$ variation in a particular 
$dN_{ch}/d\eta$ bin for different $L+R$ bins. On the other hand, in HIJING the $\la p_T\ra$ trends do not depend on the binning 
procedure and is a single-valued with $N_{part}$. We conclude that this starkly different trends of $\la p_T\ra$ in the two 
cases should be stemming from the fact that medium effects which are taken into account in AMPT are missing in HIJING. In 
the different $L+R$ bins there is different degree of medium interactions and collectivity resulting in different values of the 
final $\la p_T\ra$. Our earlier observation of different values of $v_2/\varepsilon_2$ along different $L+R$ bins support the 
above conclusion as well. Thus measurement of $\la p_T\ra$ with such combined binning procedure can also throw 
light on the degree of collectivity achieved. We note from Fig.~\ref{fig.v2bye2} that in a particular $dN_{ch}/d\eta$ bin, those $L+R$ bins which 
have the least $v_2/\varepsilon_2$ also have the highest $\la p_T\ra$ in Fig.~\ref{fig.AMPTvsHIJING} (a).

\section{Summary}
\label{sec.summary}
We have demonstrated using the AMPT model the important role played by the spectators to determine 
the initial condition in heavy-ion collisions. The standard procedure involves binning events 
by their final state multiplicity. This however puts events with varying inital conditions into the same bin as 
long as they produce similar multiplicity. We demonstrate that by further binning events according to the total number 
of spectator neutrons, it is possible to separate events with different initial conditions which were earlier 
clubbed together in the same centrality bin. This new methodology provides an opportunity to study events with rare initial 
conditions. Moreover it is possible to vary $\varepsilon_2$ and $\varepsilon_3$ independently 
of each other. This enables one to extract the contribution due to non-linear mode coupling between $v_2$ and $v_4$ and 
$\l v_2,v_3\r$ and $v_5$. It is important to note that for this purpose it is not essential to know $L+R$ 
very accurately. We found the variation of $dN_{ch}/d\eta$ with $N_{\text{part}}$ to be much different for $L+R$ bins 
compared to usual centrality bins thus allowing us to study different energy deposition mechanism within the same 
centrality. We argued that in a given centrality bin, larger $L+R$ bins have higher energy gradients and more 
number of energy hot spots as compared to smaller $L+R$ bins which result in strong inhomogenities in the inital 
conditions. This calls for larger viscosity driven effects and hence smaller $v_2/\varepsilon_2$ for bins with 
higher $L+R$. This also results in the breaking of the scaling relation between $v_2/\varepsilon_2$ and 
$\frac{1}{S}dN_{ch}/d\eta$ by the $L+R$ bins. A comparatively milder breaking of the acoustic scaling 
relation between $\ln\l v_n/\varepsilon_n\r$ and 
the initial system transverse size is observed for both centrality as well as $L+R$ bins. The results from this study 
suggest that one might be able to extract a more accurate value of the $\eta/s$ ratio with the introduction of 
the $L+R$ bins. We also observe that $\la p_T\ra$ in combined bins of $dN_{ch}/d\eta$ and $L+R$ is a good probe 
to measure the degree of medium interaction and again in this case a precise measurement of $L+R$ is not necessary. 
Hence even with the current performance of the ZDCs, we should be able to perform some of these analysis in data.

\section{Acknowledgement}
SC acknowledges ``Centre for Nuclear Theory" [PIC XII-R$\&$D-VEC-5.02.0500], Variable Energy Cyclotron Centre, 
India for support. BM acknowledges financial support by the DAE-SRC and DST-Swarnjayanti projects.

\bibliographystyle{apsrev4-1}
\bibliography{IS}

\begin{thebibliography}{31}%
\makeatletter
\providecommand \@ifxundefined [1]{%
 \@ifx{#1\undefined}
}%
\providecommand \@ifnum [1]{%
 \ifnum #1\expandafter \@firstoftwo
 \else \expandafter \@secondoftwo
 \fi
}%
\providecommand \@ifx [1]{%
 \ifx #1\expandafter \@firstoftwo
 \else \expandafter \@secondoftwo
 \fi
}%
\providecommand \natexlab [1]{#1}%
\providecommand \enquote  [1]{``#1''}%
\providecommand \bibnamefont  [1]{#1}%
\providecommand \bibfnamefont [1]{#1}%
\providecommand \citenamefont [1]{#1}%
\providecommand \href@noop [0]{\@secondoftwo}%
\providecommand \href [0]{\begingroup \@sanitize@url \@href}%
\providecommand \@href[1]{\@@startlink{#1}\@@href}%
\providecommand \@@href[1]{\endgroup#1\@@endlink}%
\providecommand \@sanitize@url [0]{\catcode `\\12\catcode `\$12\catcode
  `\&12\catcode `\#12\catcode `\^12\catcode `\_12\catcode `\%12\relax}%
\providecommand \@@startlink[1]{}%
\providecommand \@@endlink[0]{}%
\providecommand \url  [0]{\begingroup\@sanitize@url \@url }%
\providecommand \@url [1]{\endgroup\@href {#1}{\urlprefix }}%
\providecommand \urlprefix  [0]{URL }%
\providecommand \Eprint [0]{\href }%
\providecommand \doibase [0]{http://dx.doi.org/}%
\providecommand \selectlanguage [0]{\@gobble}%
\providecommand \bibinfo  [0]{\@secondoftwo}%
\providecommand \bibfield  [0]{\@secondoftwo}%
\providecommand \translation [1]{[#1]}%
\providecommand \BibitemOpen [0]{}%
\providecommand \bibitemStop [0]{}%
\providecommand \bibitemNoStop [0]{.\EOS\space}%
\providecommand \EOS [0]{\spacefactor3000\relax}%
\providecommand \BibitemShut  [1]{\csname bibitem#1\endcsname}%
\let\auto@bib@innerbib\@empty
\bibitem [{\citenamefont {Drescher}\ \emph {et~al.}(2007)\citenamefont
  {Drescher}, \citenamefont {Dumitru}, \citenamefont {Gombeaud},\ and\
  \citenamefont {Ollitrault}}]{Drescher:2007cd}%
  \BibitemOpen
  \bibfield  {author} {\bibinfo {author} {\bibfnamefont {H.-J.}\ \bibnamefont
  {Drescher}}, \bibinfo {author} {\bibfnamefont {A.}~\bibnamefont {Dumitru}},
  \bibinfo {author} {\bibfnamefont {C.}~\bibnamefont {Gombeaud}}, \ and\
  \bibinfo {author} {\bibfnamefont {J.-Y.}\ \bibnamefont {Ollitrault}},\ }\href
  {\doibase 10.1103/PhysRevC.76.024905} {\bibfield  {journal} {\bibinfo
  {journal} {Phys. Rev.}\ }\textbf {\bibinfo {volume} {C 76}},\ \bibinfo
  {pages} {024905} (\bibinfo {year} {2007})},\ \Eprint
  {http://arxiv.org/abs/0704.3553} {arXiv:0704.3553 [nucl-th]} \BibitemShut
  {NoStop}%
\bibitem [{\citenamefont {Romatschke}\ and\ \citenamefont
  {Romatschke}(2007)}]{Romatschke:2007mq}%
  \BibitemOpen
  \bibfield  {author} {\bibinfo {author} {\bibfnamefont {P.}~\bibnamefont
  {Romatschke}}\ and\ \bibinfo {author} {\bibfnamefont {U.}~\bibnamefont
  {Romatschke}},\ }\href {\doibase 10.1103/PhysRevLett.99.172301} {\bibfield
  {journal} {\bibinfo  {journal} {Phys. Rev. Lett.}\ }\textbf {\bibinfo
  {volume} {99}},\ \bibinfo {pages} {172301} (\bibinfo {year} {2007})},\
  \Eprint {http://arxiv.org/abs/0706.1522} {arXiv:0706.1522 [nucl-th]}
  \BibitemShut {NoStop}%
\bibitem [{\citenamefont {Luzum}\ and\ \citenamefont
  {Romatschke}(2008)}]{Luzum:2008cw}%
  \BibitemOpen
  \bibfield  {author} {\bibinfo {author} {\bibfnamefont {M.}~\bibnamefont
  {Luzum}}\ and\ \bibinfo {author} {\bibfnamefont {P.}~\bibnamefont
  {Romatschke}},\ }\href {\doibase 10.1103/PhysRevC.78.034915,
  10.1103/PhysRevC.79.039903} {\bibfield  {journal} {\bibinfo  {journal} {Phys.
  Rev.}\ }\textbf {\bibinfo {volume} {C 78}},\ \bibinfo {pages} {034915}
  (\bibinfo {year} {2008})},\ \bibinfo {note} {[Erratum: Phys.
  Rev.C79,039903(2009)]},\ \Eprint {http://arxiv.org/abs/0804.4015}
  {arXiv:0804.4015 [nucl-th]} \BibitemShut {NoStop}%
\bibitem [{\citenamefont {Song}\ \emph {et~al.}(2011)\citenamefont {Song},
  \citenamefont {Bass}, \citenamefont {Heinz}, \citenamefont {Hirano},\ and\
  \citenamefont {Shen}}]{Song:2010mg}%
  \BibitemOpen
  \bibfield  {author} {\bibinfo {author} {\bibfnamefont {H.}~\bibnamefont
  {Song}}, \bibinfo {author} {\bibfnamefont {S.~A.}\ \bibnamefont {Bass}},
  \bibinfo {author} {\bibfnamefont {U.}~\bibnamefont {Heinz}}, \bibinfo
  {author} {\bibfnamefont {T.}~\bibnamefont {Hirano}}, \ and\ \bibinfo {author}
  {\bibfnamefont {C.}~\bibnamefont {Shen}},\ }\href {\doibase
  10.1103/PhysRevLett.106.192301, 10.1103/PhysRevLett.109.139904} {\bibfield
  {journal} {\bibinfo  {journal} {Phys. Rev. Lett.}\ }\textbf {\bibinfo
  {volume} {106}},\ \bibinfo {pages} {192301} (\bibinfo {year} {2011})},\
  \bibinfo {note} {[Erratum: Phys. Rev. Lett.109,139904(2012)]},\ \Eprint
  {http://arxiv.org/abs/1011.2783} {arXiv:1011.2783 [nucl-th]} \BibitemShut
  {NoStop}%
\bibitem [{\citenamefont {Roy}\ \emph {et~al.}(2012)\citenamefont {Roy},
  \citenamefont {Chaudhuri},\ and\ \citenamefont {Mohanty}}]{Roy:2012jb}%
  \BibitemOpen
  \bibfield  {author} {\bibinfo {author} {\bibfnamefont {V.}~\bibnamefont
  {Roy}}, \bibinfo {author} {\bibfnamefont {A.~K.}\ \bibnamefont {Chaudhuri}},
  \ and\ \bibinfo {author} {\bibfnamefont {B.}~\bibnamefont {Mohanty}},\ }\href
  {\doibase 10.1103/PhysRevC.86.014902} {\bibfield  {journal} {\bibinfo
  {journal} {Phys. Rev.}\ }\textbf {\bibinfo {volume} {C 86}},\ \bibinfo
  {pages} {014902} (\bibinfo {year} {2012})},\ \Eprint
  {http://arxiv.org/abs/1204.2347} {arXiv:1204.2347 [nucl-th]} \BibitemShut
  {NoStop}%
\bibitem [{\citenamefont {Roy}\ \emph {et~al.}(2013)\citenamefont {Roy},
  \citenamefont {Mohanty},\ and\ \citenamefont {Chaudhuri}}]{Roy:2012pn}%
  \BibitemOpen
  \bibfield  {author} {\bibinfo {author} {\bibfnamefont {V.}~\bibnamefont
  {Roy}}, \bibinfo {author} {\bibfnamefont {B.}~\bibnamefont {Mohanty}}, \ and\
  \bibinfo {author} {\bibfnamefont {A.~K.}\ \bibnamefont {Chaudhuri}},\ }\href
  {\doibase 10.1088/0954-3899/40/6/065103} {\bibfield  {journal} {\bibinfo
  {journal} {J. Phys.}\ }\textbf {\bibinfo {volume} {G 40}},\ \bibinfo {pages}
  {065103} (\bibinfo {year} {2013})},\ \Eprint {http://arxiv.org/abs/1210.1700}
  {arXiv:1210.1700 [nucl-th]} \BibitemShut {NoStop}%
\bibitem [{\citenamefont {Chatterjee}\ and\ \citenamefont
  {Tribedy}(2015)}]{Chatterjee:2014sea}%
  \BibitemOpen
  \bibfield  {author} {\bibinfo {author} {\bibfnamefont {S.}~\bibnamefont
  {Chatterjee}}\ and\ \bibinfo {author} {\bibfnamefont {P.}~\bibnamefont
  {Tribedy}},\ }\href {\doibase 10.1103/PhysRevC.92.011902} {\bibfield
  {journal} {\bibinfo  {journal} {Phys. Rev.}\ }\textbf {\bibinfo {volume} {C
  92}},\ \bibinfo {pages} {011902} (\bibinfo {year} {2015})},\ \Eprint
  {http://arxiv.org/abs/1412.5103} {arXiv:1412.5103 [nucl-th]} \BibitemShut
  {NoStop}%
\bibitem [{\citenamefont {Bairathi}\ \emph {et~al.}(2015)\citenamefont
  {Bairathi}, \citenamefont {Haque},\ and\ \citenamefont
  {Mohanty}}]{Bairathi:2015uba}%
  \BibitemOpen
  \bibfield  {author} {\bibinfo {author} {\bibfnamefont {V.}~\bibnamefont
  {Bairathi}}, \bibinfo {author} {\bibfnamefont {M.~R.}\ \bibnamefont {Haque}},
  \ and\ \bibinfo {author} {\bibfnamefont {B.}~\bibnamefont {Mohanty}},\ }\href
  {\doibase 10.1103/PhysRevC.91.054903} {\bibfield  {journal} {\bibinfo
  {journal} {Phys. Rev.}\ }\textbf {\bibinfo {volume} {C 91}},\ \bibinfo
  {pages} {054903} (\bibinfo {year} {2015})},\ \Eprint
  {http://arxiv.org/abs/1504.04719} {arXiv:1504.04719 [nucl-ex]} \BibitemShut
  {NoStop}%
\bibitem [{\citenamefont {Jia}\ \emph {et~al.}(2015)\citenamefont {Jia},
  \citenamefont {Radhakrishnan},\ and\ \citenamefont {Zhou}}]{Jia:2015jga}%
  \BibitemOpen
  \bibfield  {author} {\bibinfo {author} {\bibfnamefont {J.}~\bibnamefont
  {Jia}}, \bibinfo {author} {\bibfnamefont {S.}~\bibnamefont {Radhakrishnan}},
  \ and\ \bibinfo {author} {\bibfnamefont {M.}~\bibnamefont {Zhou}},\
  }\href@noop {} {\  (\bibinfo {year} {2015})},\ \Eprint
  {http://arxiv.org/abs/1506.03496} {arXiv:1506.03496 [nucl-th]} \BibitemShut
  {NoStop}%
\bibitem [{\citenamefont {Abelev}\ \emph {et~al.}(2013)\citenamefont {Abelev}
  \emph {et~al.}}]{Abelev:2013qoq}%
  \BibitemOpen
  \bibfield  {author} {\bibinfo {author} {\bibfnamefont {B.}~\bibnamefont
  {Abelev}} \emph {et~al.} (\bibinfo {collaboration} {ALICE}),\ }\href
  {\doibase 10.1103/PhysRevC.88.044909} {\bibfield  {journal} {\bibinfo
  {journal} {Phys. Rev.}\ }\textbf {\bibinfo {volume} {C88}},\ \bibinfo {pages}
  {044909} (\bibinfo {year} {2013})},\ \Eprint {http://arxiv.org/abs/1301.4361}
  {arXiv:1301.4361 [nucl-ex]} \BibitemShut {NoStop}%
\bibitem [{\citenamefont {Tarafdar}\ \emph {et~al.}(2014)\citenamefont
  {Tarafdar}, \citenamefont {Citron},\ and\ \citenamefont
  {Milov}}]{Tarafdar:2014oua}%
  \BibitemOpen
  \bibfield  {author} {\bibinfo {author} {\bibfnamefont {S.}~\bibnamefont
  {Tarafdar}}, \bibinfo {author} {\bibfnamefont {Z.}~\bibnamefont {Citron}}, \
  and\ \bibinfo {author} {\bibfnamefont {A.}~\bibnamefont {Milov}},\ }\href
  {\doibase 10.1016/j.nima.2014.09.060} {\bibfield  {journal} {\bibinfo
  {journal} {Nucl. Instrum. Meth.}\ }\textbf {\bibinfo {volume} {A768}},\
  \bibinfo {pages} {170} (\bibinfo {year} {2014})},\ \Eprint
  {http://arxiv.org/abs/1405.4555} {arXiv:1405.4555 [nucl-ex]} \BibitemShut
  {NoStop}%
\bibitem [{\citenamefont {Lin}\ and\ \citenamefont {Ko}(2002)}]{Lin:2001zk}%
  \BibitemOpen
  \bibfield  {author} {\bibinfo {author} {\bibfnamefont {Z.-w.}\ \bibnamefont
  {Lin}}\ and\ \bibinfo {author} {\bibfnamefont {C.}~\bibnamefont {Ko}},\
  }\href {\doibase 10.1103/PhysRevC.65.034904} {\bibfield  {journal} {\bibinfo
  {journal} {Phys. Rev.}\ }\textbf {\bibinfo {volume} {C 65}},\ \bibinfo
  {pages} {034904} (\bibinfo {year} {2002})},\ \Eprint
  {http://arxiv.org/abs/nucl-th/0108039} {arXiv:nucl-th/0108039 [nucl-th]}
  \BibitemShut {NoStop}%
\bibitem [{\citenamefont {Lin}\ \emph {et~al.}(2005)\citenamefont {Lin},
  \citenamefont {Ko}, \citenamefont {Li}, \citenamefont {Zhang},\ and\
  \citenamefont {Pal}}]{Lin:2004en}%
  \BibitemOpen
  \bibfield  {author} {\bibinfo {author} {\bibfnamefont {Z.-W.}\ \bibnamefont
  {Lin}}, \bibinfo {author} {\bibfnamefont {C.~M.}\ \bibnamefont {Ko}},
  \bibinfo {author} {\bibfnamefont {B.-A.}\ \bibnamefont {Li}}, \bibinfo
  {author} {\bibfnamefont {B.}~\bibnamefont {Zhang}}, \ and\ \bibinfo {author}
  {\bibfnamefont {S.}~\bibnamefont {Pal}},\ }\href {\doibase
  10.1103/PhysRevC.72.064901} {\bibfield  {journal} {\bibinfo  {journal} {Phys.
  Rev.}\ }\textbf {\bibinfo {volume} {C 72}},\ \bibinfo {pages} {064901}
  (\bibinfo {year} {2005})},\ \Eprint {http://arxiv.org/abs/nucl-th/0411110}
  {arXiv:nucl-th/0411110 [nucl-th]} \BibitemShut {NoStop}%
\bibitem [{\citenamefont {Wang}\ and\ \citenamefont
  {Gyulassy}(1991)}]{Wang:1991hta}%
  \BibitemOpen
  \bibfield  {author} {\bibinfo {author} {\bibfnamefont {X.-N.}\ \bibnamefont
  {Wang}}\ and\ \bibinfo {author} {\bibfnamefont {M.}~\bibnamefont
  {Gyulassy}},\ }\href {\doibase 10.1103/PhysRevD.44.3501} {\bibfield
  {journal} {\bibinfo  {journal} {Phys. Rev.}\ }\textbf {\bibinfo {volume} {D
  44}},\ \bibinfo {pages} {3501} (\bibinfo {year} {1991})}\BibitemShut
  {NoStop}%
\bibitem [{\citenamefont {Zhang}(1998)}]{Zhang:1997ej}%
  \BibitemOpen
  \bibfield  {author} {\bibinfo {author} {\bibfnamefont {B.}~\bibnamefont
  {Zhang}},\ }\href {\doibase 10.1016/S0010-4655(98)00010-1} {\bibfield
  {journal} {\bibinfo  {journal} {Comput. Phys. Commun.}\ }\textbf {\bibinfo
  {volume} {109}},\ \bibinfo {pages} {193} (\bibinfo {year} {1998})},\ \Eprint
  {http://arxiv.org/abs/nucl-th/9709009} {arXiv:nucl-th/9709009 [nucl-th]}
  \BibitemShut {NoStop}%
\bibitem [{\citenamefont {Andersson}\ \emph {et~al.}(1983)\citenamefont
  {Andersson}, \citenamefont {Gustafson}, \citenamefont {Ingelman},\ and\
  \citenamefont {Sjostrand}}]{Andersson:1983ia}%
  \BibitemOpen
  \bibfield  {author} {\bibinfo {author} {\bibfnamefont {B.}~\bibnamefont
  {Andersson}}, \bibinfo {author} {\bibfnamefont {G.}~\bibnamefont
  {Gustafson}}, \bibinfo {author} {\bibfnamefont {G.}~\bibnamefont {Ingelman}},
  \ and\ \bibinfo {author} {\bibfnamefont {T.}~\bibnamefont {Sjostrand}},\
  }\href {\doibase 10.1016/0370-1573(83)90080-7} {\bibfield  {journal}
  {\bibinfo  {journal} {Phys. Rept.}\ }\textbf {\bibinfo {volume} {97}},\
  \bibinfo {pages} {31} (\bibinfo {year} {1983})}\BibitemShut {NoStop}%
\bibitem [{\citenamefont {Bhalerao}\ and\ \citenamefont
  {Ollitrault}(2006)}]{Bhalerao:2006tp}%
  \BibitemOpen
  \bibfield  {author} {\bibinfo {author} {\bibfnamefont {R.~S.}\ \bibnamefont
  {Bhalerao}}\ and\ \bibinfo {author} {\bibfnamefont {J.-Y.}\ \bibnamefont
  {Ollitrault}},\ }\href {\doibase 10.1016/j.physletb.2006.08.055} {\bibfield
  {journal} {\bibinfo  {journal} {Phys. Lett.}\ }\textbf {\bibinfo {volume} {B
  641}},\ \bibinfo {pages} {260} (\bibinfo {year} {2006})},\ \Eprint
  {http://arxiv.org/abs/nucl-th/0607009} {arXiv:nucl-th/0607009 [nucl-th]}
  \BibitemShut {NoStop}%
\bibitem [{\citenamefont {Bialas}\ \emph {et~al.}(1976)\citenamefont {Bialas},
  \citenamefont {Bleszynski},\ and\ \citenamefont {Czyz}}]{Bialas:1976ed}%
  \BibitemOpen
  \bibfield  {author} {\bibinfo {author} {\bibfnamefont {A.}~\bibnamefont
  {Bialas}}, \bibinfo {author} {\bibfnamefont {M.}~\bibnamefont {Bleszynski}},
  \ and\ \bibinfo {author} {\bibfnamefont {W.}~\bibnamefont {Czyz}},\ }\href
  {\doibase 10.1016/0550-3213(76)90329-1} {\bibfield  {journal} {\bibinfo
  {journal} {Nucl. Phys.}\ }\textbf {\bibinfo {volume} {B 111}},\ \bibinfo
  {pages} {461} (\bibinfo {year} {1976})}\BibitemShut {NoStop}%
\bibitem [{\citenamefont {Wang}\ and\ \citenamefont
  {Gyulassy}(2001)}]{Wang:2000bf}%
  \BibitemOpen
  \bibfield  {author} {\bibinfo {author} {\bibfnamefont {X.-N.}\ \bibnamefont
  {Wang}}\ and\ \bibinfo {author} {\bibfnamefont {M.}~\bibnamefont
  {Gyulassy}},\ }\href {\doibase 10.1103/PhysRevLett.86.3496} {\bibfield
  {journal} {\bibinfo  {journal} {Phys. Rev. Lett.}\ }\textbf {\bibinfo
  {volume} {86}},\ \bibinfo {pages} {3496} (\bibinfo {year} {2001})},\ \Eprint
  {http://arxiv.org/abs/nucl-th/0008014} {arXiv:nucl-th/0008014 [nucl-th]}
  \BibitemShut {NoStop}%
\bibitem [{\citenamefont {Kharzeev}\ and\ \citenamefont
  {Nardi}(2001)}]{Kharzeev:2000ph}%
  \BibitemOpen
  \bibfield  {author} {\bibinfo {author} {\bibfnamefont {D.}~\bibnamefont
  {Kharzeev}}\ and\ \bibinfo {author} {\bibfnamefont {M.}~\bibnamefont
  {Nardi}},\ }\href {\doibase 10.1016/S0370-2693(01)00457-9} {\bibfield
  {journal} {\bibinfo  {journal} {Phys. Lett.}\ }\textbf {\bibinfo {volume} {B
  507}},\ \bibinfo {pages} {121} (\bibinfo {year} {2001})},\ \Eprint
  {http://arxiv.org/abs/nucl-th/0012025} {arXiv:nucl-th/0012025 [nucl-th]}
  \BibitemShut {NoStop}%
\bibitem [{\citenamefont {Poskanzer}\ and\ \citenamefont
  {Voloshin}(1998)}]{Poskanzer:1998yz}%
  \BibitemOpen
  \bibfield  {author} {\bibinfo {author} {\bibfnamefont {A.~M.}\ \bibnamefont
  {Poskanzer}}\ and\ \bibinfo {author} {\bibfnamefont {S.~A.}\ \bibnamefont
  {Voloshin}},\ }\href {\doibase 10.1103/PhysRevC.58.1671} {\bibfield
  {journal} {\bibinfo  {journal} {Phys. Rev.}\ }\textbf {\bibinfo {volume} {C
  58}},\ \bibinfo {pages} {1671} (\bibinfo {year} {1998})},\ \Eprint
  {http://arxiv.org/abs/nucl-ex/9805001} {arXiv:nucl-ex/9805001 [nucl-ex]}
  \BibitemShut {NoStop}%
\bibitem [{\citenamefont {Schukraft}\ \emph {et~al.}(2013)\citenamefont
  {Schukraft}, \citenamefont {Timmins},\ and\ \citenamefont
  {Voloshin}}]{Schukraft:2012ah}%
  \BibitemOpen
  \bibfield  {author} {\bibinfo {author} {\bibfnamefont {J.}~\bibnamefont
  {Schukraft}}, \bibinfo {author} {\bibfnamefont {A.}~\bibnamefont {Timmins}},
  \ and\ \bibinfo {author} {\bibfnamefont {S.~A.}\ \bibnamefont {Voloshin}},\
  }\href {\doibase 10.1016/j.physletb.2013.01.045} {\bibfield  {journal}
  {\bibinfo  {journal} {Phys. Lett.}\ }\textbf {\bibinfo {volume} {B 719}},\
  \bibinfo {pages} {394} (\bibinfo {year} {2013})},\ \Eprint
  {http://arxiv.org/abs/1208.4563} {arXiv:1208.4563 [nucl-ex]} \BibitemShut
  {NoStop}%
\bibitem [{\citenamefont {Huo}\ \emph {et~al.}(2014)\citenamefont {Huo},
  \citenamefont {Jia},\ and\ \citenamefont {Mohapatra}}]{Huo:2013qma}%
  \BibitemOpen
  \bibfield  {author} {\bibinfo {author} {\bibfnamefont {P.}~\bibnamefont
  {Huo}}, \bibinfo {author} {\bibfnamefont {J.}~\bibnamefont {Jia}}, \ and\
  \bibinfo {author} {\bibfnamefont {S.}~\bibnamefont {Mohapatra}},\ }\href
  {\doibase 10.1103/PhysRevC.90.024910} {\bibfield  {journal} {\bibinfo
  {journal} {Phys. Rev.}\ }\textbf {\bibinfo {volume} {C 90}},\ \bibinfo
  {pages} {024910} (\bibinfo {year} {2014})},\ \Eprint
  {http://arxiv.org/abs/1311.7091} {arXiv:1311.7091 [nucl-ex]} \BibitemShut
  {NoStop}%
\bibitem [{\citenamefont {Bhalerao}\ \emph {et~al.}(2005)\citenamefont
  {Bhalerao}, \citenamefont {Blaizot}, \citenamefont {Borghini},\ and\
  \citenamefont {Ollitrault}}]{Bhalerao:2005mm}%
  \BibitemOpen
  \bibfield  {author} {\bibinfo {author} {\bibfnamefont {R.~S.}\ \bibnamefont
  {Bhalerao}}, \bibinfo {author} {\bibfnamefont {J.-P.}\ \bibnamefont
  {Blaizot}}, \bibinfo {author} {\bibfnamefont {N.}~\bibnamefont {Borghini}}, \
  and\ \bibinfo {author} {\bibfnamefont {J.-Y.}\ \bibnamefont {Ollitrault}},\
  }\href {\doibase 10.1016/j.physletb.2005.08.131} {\bibfield  {journal}
  {\bibinfo  {journal} {Phys. Lett.}\ }\textbf {\bibinfo {volume} {B 627}},\
  \bibinfo {pages} {49} (\bibinfo {year} {2005})},\ \Eprint
  {http://arxiv.org/abs/nucl-th/0508009} {arXiv:nucl-th/0508009 [nucl-th]}
  \BibitemShut {NoStop}%
\bibitem [{\citenamefont {Heiselberg}\ and\ \citenamefont
  {Levy}(1999)}]{Heiselberg:1998es}%
  \BibitemOpen
  \bibfield  {author} {\bibinfo {author} {\bibfnamefont {H.}~\bibnamefont
  {Heiselberg}}\ and\ \bibinfo {author} {\bibfnamefont {A.-M.}\ \bibnamefont
  {Levy}},\ }\href {\doibase 10.1103/PhysRevC.59.2716} {\bibfield  {journal}
  {\bibinfo  {journal} {Phys. Rev.}\ }\textbf {\bibinfo {volume} {C 59}},\
  \bibinfo {pages} {2716} (\bibinfo {year} {1999})},\ \Eprint
  {http://arxiv.org/abs/nucl-th/9812034} {arXiv:nucl-th/9812034 [nucl-th]}
  \BibitemShut {NoStop}%
\bibitem [{\citenamefont {Voloshin}\ and\ \citenamefont
  {Poskanzer}(2000)}]{Voloshin:1999gs}%
  \BibitemOpen
  \bibfield  {author} {\bibinfo {author} {\bibfnamefont {S.~A.}\ \bibnamefont
  {Voloshin}}\ and\ \bibinfo {author} {\bibfnamefont {A.~M.}\ \bibnamefont
  {Poskanzer}},\ }\href {\doibase 10.1016/S0370-2693(00)00017-4} {\bibfield
  {journal} {\bibinfo  {journal} {Phys. Lett.}\ }\textbf {\bibinfo {volume} {B
  474}},\ \bibinfo {pages} {27} (\bibinfo {year} {2000})},\ \Eprint
  {http://arxiv.org/abs/nucl-th/9906075} {arXiv:nucl-th/9906075 [nucl-th]}
  \BibitemShut {NoStop}%
\bibitem [{\citenamefont {Kolb}\ \emph {et~al.}(2001)\citenamefont {Kolb},
  \citenamefont {Huovinen}, \citenamefont {Heinz},\ and\ \citenamefont
  {Heiselberg}}]{Kolb:2000fha}%
  \BibitemOpen
  \bibfield  {author} {\bibinfo {author} {\bibfnamefont {P.~F.}\ \bibnamefont
  {Kolb}}, \bibinfo {author} {\bibfnamefont {P.}~\bibnamefont {Huovinen}},
  \bibinfo {author} {\bibfnamefont {U.~W.}\ \bibnamefont {Heinz}}, \ and\
  \bibinfo {author} {\bibfnamefont {H.}~\bibnamefont {Heiselberg}},\ }\href
  {\doibase 10.1016/S0370-2693(01)00079-X} {\bibfield  {journal} {\bibinfo
  {journal} {Phys. Lett.}\ }\textbf {\bibinfo {volume} {B 500}},\ \bibinfo
  {pages} {232} (\bibinfo {year} {2001})},\ \Eprint
  {http://arxiv.org/abs/hep-ph/0012137} {arXiv:hep-ph/0012137 [hep-ph]}
  \BibitemShut {NoStop}%
\bibitem [{\citenamefont {Staig}\ and\ \citenamefont
  {Shuryak}(2011)}]{Staig:2010pn}%
  \BibitemOpen
  \bibfield  {author} {\bibinfo {author} {\bibfnamefont {P.}~\bibnamefont
  {Staig}}\ and\ \bibinfo {author} {\bibfnamefont {E.}~\bibnamefont
  {Shuryak}},\ }\href {\doibase 10.1103/PhysRevC.84.034908} {\bibfield
  {journal} {\bibinfo  {journal} {Phys. Rev.}\ }\textbf {\bibinfo {volume} {C
  84}},\ \bibinfo {pages} {034908} (\bibinfo {year} {2011})},\ \Eprint
  {http://arxiv.org/abs/1008.3139} {arXiv:1008.3139 [nucl-th]} \BibitemShut
  {NoStop}%
\bibitem [{\citenamefont {Lacey}\ \emph {et~al.}(2011)\citenamefont {Lacey},
  \citenamefont {Taranenko}, \citenamefont {Ajitanand},\ and\ \citenamefont
  {Alexander}}]{Lacey:2011ug}%
  \BibitemOpen
  \bibfield  {author} {\bibinfo {author} {\bibfnamefont {R.~A.}\ \bibnamefont
  {Lacey}}, \bibinfo {author} {\bibfnamefont {A.}~\bibnamefont {Taranenko}},
  \bibinfo {author} {\bibfnamefont {N.~N.}\ \bibnamefont {Ajitanand}}, \ and\
  \bibinfo {author} {\bibfnamefont {J.~M.}\ \bibnamefont {Alexander}},\
  }\href@noop {} {\  (\bibinfo {year} {2011})},\ \Eprint
  {http://arxiv.org/abs/1105.3782} {arXiv:1105.3782 [nucl-ex]} \BibitemShut
  {NoStop}%
\bibitem [{\citenamefont {Lacey}\ \emph {et~al.}(2013)\citenamefont {Lacey},
  \citenamefont {Gu}, \citenamefont {Gong}, \citenamefont {Reynolds},
  \citenamefont {Ajitanand} \emph {et~al.}}]{Lacey:2013is}%
  \BibitemOpen
  \bibfield  {author} {\bibinfo {author} {\bibfnamefont {R.~A.}\ \bibnamefont
  {Lacey}}, \bibinfo {author} {\bibfnamefont {Y.}~\bibnamefont {Gu}}, \bibinfo
  {author} {\bibfnamefont {X.}~\bibnamefont {Gong}}, \bibinfo {author}
  {\bibfnamefont {D.}~\bibnamefont {Reynolds}}, \bibinfo {author}
  {\bibfnamefont {N.}~\bibnamefont {Ajitanand}},  \emph {et~al.},\ }\href@noop
  {} {\  (\bibinfo {year} {2013})},\ \Eprint {http://arxiv.org/abs/1301.0165}
  {arXiv:1301.0165} \BibitemShut {NoStop}%
\bibitem [{\citenamefont {Lacey}\ \emph {et~al.}(2014)\citenamefont {Lacey},
  \citenamefont {Taranenko}, \citenamefont {Jia}, \citenamefont {Reynolds},
  \citenamefont {Ajitanand} \emph {et~al.}}]{Lacey:2013qua}%
  \BibitemOpen
  \bibfield  {author} {\bibinfo {author} {\bibfnamefont {R.~A.}\ \bibnamefont
  {Lacey}}, \bibinfo {author} {\bibfnamefont {A.}~\bibnamefont {Taranenko}},
  \bibinfo {author} {\bibfnamefont {J.}~\bibnamefont {Jia}}, \bibinfo {author}
  {\bibfnamefont {D.}~\bibnamefont {Reynolds}}, \bibinfo {author}
  {\bibfnamefont {N.}~\bibnamefont {Ajitanand}},  \emph {et~al.},\ }\href
  {\doibase 10.1103/PhysRevLett.112.082302} {\bibfield  {journal} {\bibinfo
  {journal} {Phys. Rev. Lett.}\ }\textbf {\bibinfo {volume} {112}},\ \bibinfo
  {pages} {082302} (\bibinfo {year} {2014})},\ \Eprint
  {http://arxiv.org/abs/1305.3341} {arXiv:1305.3341 [nucl-ex]} \BibitemShut
  {NoStop}%
\end{thebibliography}%

\end{document}